\documentclass{aa}  

\usepackage{graphicx}
\usepackage{txfonts}
\usepackage{lipsum}
\usepackage{placeins}           
\usepackage{subcaption}                  
\usepackage[colorlinks=true, linkcolor=blue, citecolor=blue, filecolor=blue, urlcolor=blue]{hyperref}
\usepackage{natbib}
\bibpunct{(}{)}{;}{a}{}{,}

\begin{document}

   \title{The case of NGC\,5824, a cluster possibly embedded in a dark matter halo}
   \subtitle{}
   
   \authorrunning{D\'iaz et al.}

   \author{Paula B. D\'iaz\inst{\ref{UCH}}
          \and
          Berenice Muruaga\inst{\ref{UCH}}
          \and
          Ricardo R. Muñoz\inst{\ref{UCH}}
          \and
          Julio A. Carballo-Bello\inst{\ref{UTA}}
          \and
          Pete B. Kuzma\inst{\ref{UEd}}
          \and 
          Valentina Su\'arez\inst{\ref{UCH}}
          }

   \institute{Departamento de Astronom\'{\i}a, Universidad de Chile, Camino del Observatorio 1515, Las Condes, Santiago, Chile\\
            \email{pdiaz@das.uchile.cl} \label{UCH}
            \and
            {Instituto de Alta Investigaci\'on, Universidad de Tarapac\'a, Casilla 7D, Arica, Chile} \label{UTA}
            \and
            {Institute for Astronomy, University of Edimburgh, Royal Observatory, Blackford Hill, Edimburgh, EH9 3HJ, UK}\label{UEd}
            }

   \date{Received XXX; accepted YYY}

  \abstract
   {The globular cluster NGC\,5824 exhibits a diffuse stellar envelope that extends beyond its nominal King tidal radius and symmetrically surrounds the cluster. The origin of these stars and whether they remain gravitationally bound to the cluster center is unclear. A possible explanation is that such clusters are embedded within dark matter halos, which influences their kinematic and photometric properties. Specifically, their outer volume density profile would be characterized by a power law with an index $\gamma > -3$.}
   {In this study we assessed this photometric prediction by comparing it to the cluster's observed profile through an analysis of deep $g$-band photometry from MegaCam and DECam, combined with \textit{Gaia} DR3 proper motions and photometry.} 
   {We determined star membership using color-magnitude diagrams and proper motion constraints in order to fit King and power-law profiles to the observed profile. Additionally, we analyzed NGC\,5824's luminosity function to assess its spatial symmetry and extent.}
   {Our results show that NGC\,5824 is symmetrically extended to at least $\sim20'$\, with an outer surface density profile characterized by a power-law index of $\gamma \sim - 2.6\pm0.1$, which is consistent with the predicted values for a cluster embedded within a dark matter halo.}
   {Spectroscopic observations carried out to study the velocity dispersion profile will provide a more definitive answer regarding the dark matter content of NGC\,5824.}

   \keywords{globular clusters: general -- globular clusters: individual (NGC\,5824) -- Galaxy: halo -- dark matter.}

   \maketitle
   \nolinenumbers

\section{Introduction} \label{sec:intro}

The Milky Way (MW) galaxy is surrounded by numerous self-gravitating stellar systems orbiting within its halo. These satellites are often categorized into two distinct groups: globular clusters (GCs) and dwarf galaxies. The main distinction between these two families is their dark matter (DM) content: dwarf galaxies are considered to be DM-dominated \citep[e.g.,][]{Willman_2012}, and GCs are considered to be devoid of DM \citep[e.g.,][]{Bradford_2011,Ibata_2013}.

Milky Way GCs are old (with ages around $12$\,Gyr), luminous (visual magnitudes $M_V \sim -7$), typically populated by tens of thousands to millions of stars, have very high stellar densities \citep{Harris_1996, Bahramian_2013}, and are characterized as compact objects with half-light radii of a few tens of parsecs (see \citealt{Gratton_2019} for a review of GCs). Given the small spread in age and metallicity of their stars, they are often considered to be close approximations of simple stellar populations, although there is a great deal of evidence showing otherwise (see the review by \citealt{Milone_Milano_2022}).

Due to their ages, GCs serve as key witnesses to early galaxy formation and evolution, acting as valuable tracers of the structure of galaxies, including the MW \citep[e.g.,][]{Searle_1978, Forbes_2010, Massari_2019, Kruijssen_2020, Massari_2023}. However, there is still no clear consensus regarding their formation mechanism. For example, \cite{Peebles_1984} first proposed that GCs could be formed within DM mini-halos, while alternative formation scenarios that do not involve DM have also been suggested \citep[e.g.,][]{Conroy_2011,Naoz_2014,Lake_2023, Lake_2025}. In fact, recent cosmological simulations of realistic mechanisms have shown that both scenarios are possible \citep[][]{Trenti_2015, Kimm_2016, Ricotti_2016, Keller_2020, Ma_2020}, initiating a debate on the possibility that GCs originate via multiple formation channels. Observationally, one of the main reasons why the understanding of GC formation remains elusive is that most of the consequences of these formation channels are exhibited in the outskirts of the GCs, where stellar densities are extremely low and it therefore becomes difficult to distinguish between bona fide GC stars and MW foreground and background stars.
\begin{figure}[t]
    \centering
    \includegraphics[width=\columnwidth]{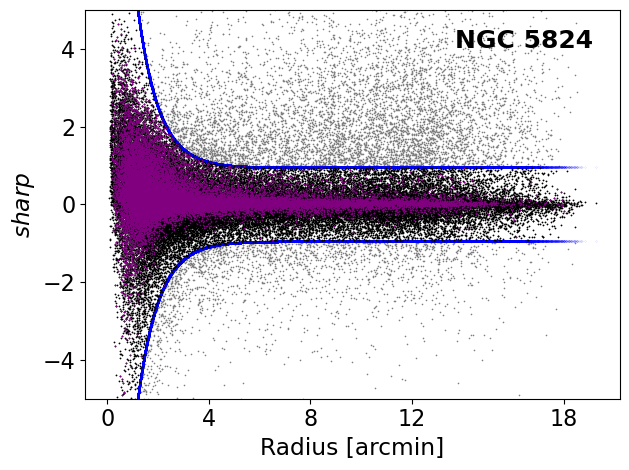}

    \vspace{0.5em}

    \includegraphics[width=\columnwidth]{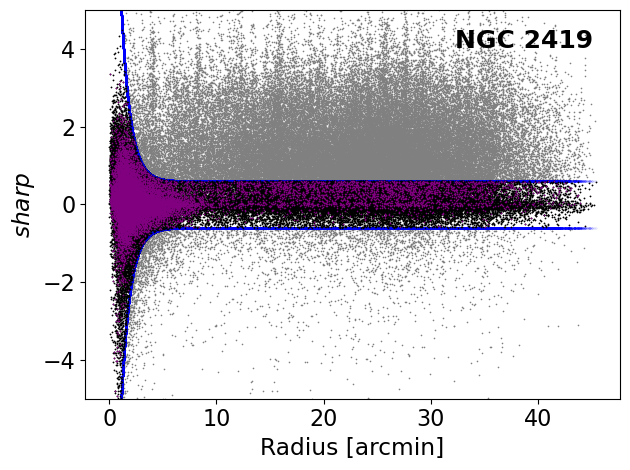}

    \caption{\textit{sharp} as a function of radius for NGC\,5824 (top panel) and NGC\,2419 (bottom panel). Gray dots represent all the detections within the MegaCam field. Blue dots indicate upward and downward exponentials applied as the \textit{sharp} filter. Black dots denote all the detections that survived the \textit{sharp} filtering. Purple dots highlight the stars selected for Test 1, which include MS, red giant, and BHB stars.}
    \label{fig:Sharp}
\end{figure}

For some time now, it has been reported that some of the outer halo GCs of the MW possess stars located beyond their King tidal radius ($r_t$), i.e., the observable radius beyond which stars are gravitationally unbound to the cluster \citep[e.g.,][]{Carballo-Bello_2011,Sanna_2012,Marino_2014,Bellazzini_2015,Kuzma_2018,Zhang_2022,Yang_2022, Kuzma_2025}. These stellar populations, when present, are generally attributed to one of two phenomena: (1) the formation of tidal tails, which are symmetric streams originating from the cluster due to the tidal pull of the MW  (see, e.g., \citealt{Odenkirchen_2001} for Palomar 5), or (2) they correspond to a diffuse stellar envelope surrounding the cluster symmetrically at distances greater than expected based on their King tidal radii (see, e.g., \citealt{Kuzma_2016} for NGC\,7089). In the latter case, it is currently unclear where these stars originate and how they manage to remain at the periphery of the cluster. A possible explanation involves GCs forming within DM halos, which allows DM to prevent stars from escaping the system's self-gravitational potential. This scenario has been explored over the years by other authors, and models of GCs with DM halos have been proposed \citep[e.g.,][]{Moore_1996, Mashchenko_2005b, Ibata_2013, Vitral_2022, Carlberg_2022}. In particular, \citet{Penarrubia_2017} predict that the presence of DM imprints discernible effects on the kinematics and photometric properties of the GCs. Specifically, clusters with higher DM content tend to be more extended, with an inflated radial velocity dispersion profile at large radii and an outer density profile characterized by a power law with index $\gamma > -3$, which asymptotically approaches $-3$ as the DM mass increases.

In this vein, one interesting case study of outer halo GCs is NGC\,5824. This cluster, first discovered in 1925 \citep{Innes_1925}, is located $32.1$\,kpc from the Sun; it is the second brightest of the outer halo clusters, with $M_V = -9.3 \pm 0.04$ \citep{Munoz_2018a}, and has a mass of $\approx 10^{6}$ $M_{\sun}$ \citep{Yuan_2022}. One particular point of interest is the fact that its density profile is not well described by a King profile \citep{Carballo-Bello_2011} but instead seems well fitted by a power law \citep{Sanna_2014}. In addition, several studies report a large extension \citep{Kuzma_2018, Yang_2022}.  \citet{Munoz_2018b} determined that the spatial extent of NGC\,5824 exceeds 30 times its Sérsic effective radius ($r_{e,s}$) and that NGC\,5824 has the highest concentration parameter ($r_{\rm t}/r_{\rm c} = 172$, with $r_c$ being its core radius) of the whole sample of the MW outer halo GCs, despite having a relatively small effective radius.

In this study we focused on testing the predictions of the Pe\~narrubia model for the outer slopes of the density profile of NGC\,5824 by comparing it with the observed profile based on deep photometry and proper motion data. In addition, we analyzed the luminosity function (LF) of NGC\,5824 to determine whether there is a diffuse stellar envelope and thus whether the primary assumption of the \citet{Penarrubia_2017} model is satisfied. We used the public photometry obtained by \citet{Munoz_2018a} and the \textit{Gaia} Data Release 3 (DR3) catalog, and supplementary Dark Energy Camera (DECam) photometry from \citet{Kuzma_2018}. Furthermore, we contrasted the density profile behavior of NGC\,5824 with that of NGC\,2419, the largest and brightest of the outer halo clusters with $M_V = -9.35 \pm 0.03$ and $r_{e,s}\sim25.7 \pm 0.2$\,pc, $r_t = 227$\,pc \citep{Munoz_2018a}, to provide a reference point for a cluster with a lower concentration parameter ($r_{\rm t}/r_{\rm c} = 50$). The contrast between these two clusters is evident not only due to their significantly different concentration parameters, but also because they represent two extremes in terms of the Sérsic index and exhibit clearly different central surface brightnesses. Specifically, NGC\,5824 has a S\'ersic index of $n=3.82\pm0.05$ and $\mu_{V,0}=11.15\pm0.08$\,$\rm{mag\,arcsec^{-2}}$, whereas NGC\,2419 has $n = 1.71\pm0.02$ and $\mu_{V,0} = 18.83 \pm 0.05$\,$\rm{mag\,arcsec^{-2}}$ \citep{Marchi-Lasch_2019}. 

The paper is organized as follows. In Sect. \ref{obs_data_red} we summarize the data acquisition and data reduction process presented by \citet{Munoz_2018a}. Section \ref{methods} describes our methodology, which involves using multiple independent datasets to construct density profiles and fit King and power-law models for comparison with the prediction of \citet{Penarrubia_2017}. Additionally, we analyze LFs to assess the cluster's spatial symmetry and extent. Our results are presented in Sect. \ref{results}. In Sect. \ref{analysis} we discuss the implications of our findings by comparing them with the literature, and finally in Sect. \ref{conclusion} we present our conclusions.

\begin{figure}[h]
    \centering
    \includegraphics[width=\columnwidth]{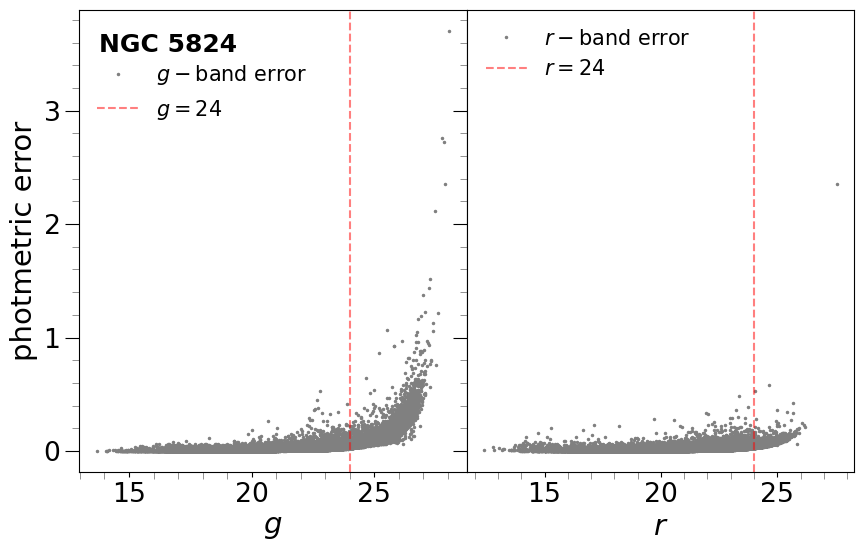}
    
    \vspace{0.1em}

    \includegraphics[width=\columnwidth]{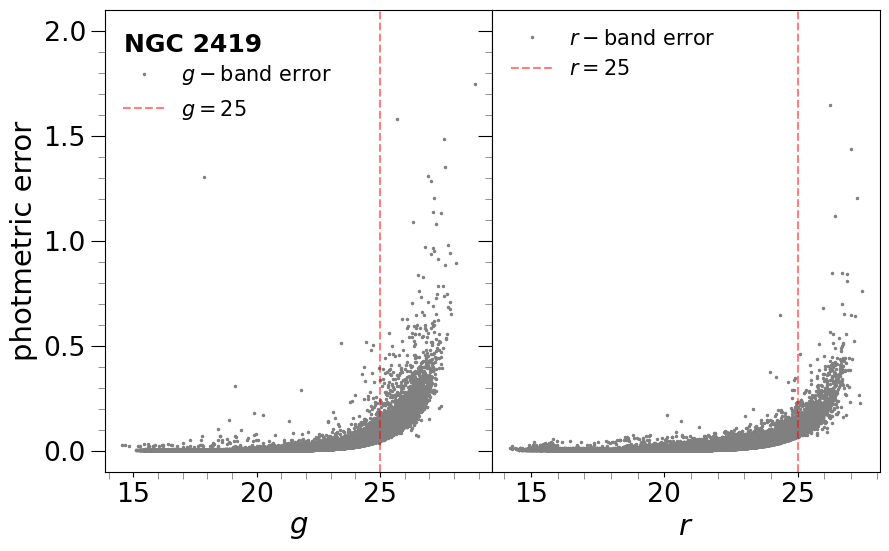}
    \caption{Photometric errors as a function of MegaCam magnitudes for NGC\,5824 (top panels) and NGC\,2419 (bottom panels). Left panels show the $g$ band and right panels the $r$ band. The dashed vertical red line indicates magnitude $24$ for NGC\,5824 and $25$ for NGC\,2419 in both bands.}
    \label{fig:photometric_errors}
\end{figure}

\section{Data and reduction} \label{obs_data_red}

In this study we analyzed the photometric and proper motion data of the GC NGC\,5824 ($\alpha_0=$\,$225.9943\,\rm{deg}$, $\delta_0=$\,$-33.0685\,\rm{deg}$) and compared them with the predictions made by the \citet{Penarrubia_2017} model. In addition, as a control case, we analyzed the photometric properties of NGC\,2419 ($\alpha_0=$\,$114.5354\,\rm{deg}$, $\delta_0=$\,$+38.8819\,\rm{deg}$), the brightest and one of the most distant outer halo clusters. The data used for both clusters come from the survey of the MW outer halo satellites carried out by \citet{Munoz_2018b}. This survey used the MegaCam imager on both the Magellan Clay telescope and the Canada-France-Hawaii Telescope (CFHT), providing deep imaging in the $g$ and $r$ bands over a wide field. The typical limiting magnitudes for point sources at a confidence level of $5\sigma$ are $g_{lim} \simeq 25.6$ and $r_{lim} \simeq 25.3$ in AB magnitudes. On the other hand, NGC\,2419 was observed with CFHT. The MegaCam on Clay covers an area of $0.4\,\rm{deg}\times0.4\,\rm{deg}$ on the sky, while on CFHT it covers $1\,\rm{deg}\times 1\,\rm{deg}$.

The MegaCam data were preprocessed following the procedure specified in \citet{Munoz_2018b}, which can be briefly described as follows: The initial astrometric and photometric solutions were refined. For the astrometric calibration of the GCs in this paper, the solutions were improved using the \textit{Gaia} DR1 catalog, achieving root-mean-square uncertainties of $\sim 0\farcs04$--$0\farcs06$. Point-spread function photometry was then carried out using DAOPHOT/ALLSTAR and ALLFRAME packages. Finally, photometric calibration was performed by determining zero points and color terms using Sloan Digital Sky Survey objects that overlapped with the survey, either as direct calibrators or as secondary standards.

Additionally, to clean the catalog of unresolved background galaxies and artifacts, we carried out a cut in the DAOPHOT morphological parameter \textit{sharp}. This parameter is a measure of how much broader or narrower an object's light profile is compared to that of a point source. In essence, \textit{sharp} reflects how widespread or narrow the point spread function is. 
We applied cuts following upward and downward exponential functions, as shown in the top and bottom panels of Fig. \ref{fig:Sharp} for NGC\,5824 and NGC\,2419, respectively.

In the case of NGC\,5824, to improve our ability to detect member stars extending to the outskirts of the cluster, we complemented these data with $G$- and $G_{rp}$-band photometry and proper motions from the \textit{Gaia} DR3 archive\footnote{\url{https://gea.esac.esa.int/archive/}} \citep{gaia_2023}. This additional data cover a circular region of $60'$ from the cluster center, with a limiting magnitude of $G_{lim} = 20.7$ mag \citep{gaia_edr3}. We cleaned this catalog by selecting objects with a probability of being a single star, but not a white dwarf, greater than $0.8$, according to the \textit{Gaia} DR3 Data Source Classification System \citep{Delchambre_2023}. In addition, we incorporated DECam photometry in the $g$ and $i$ band from \cite{Kuzma_2018}, who traced NGC\,5824 stars out to $230$\,pc by analyzing its 2D density map. For this dataset, we restricted the covering region up to $35'$ from the cluster center.

\section{Methods} \label{methods}
Given that in the outer regions of the cluster the stellar densities are much lower, making it difficult to assess the real extent of the cluster stars, we computed the radial density profiles through four different tests. Each test corresponds to a different approach for identifying cluster stars, from which radial surface density profiles were constructed to perform King and power-law fits. 

The first two tests focus on main sequence (MS), red giant branch (RGB), and blue horizontal branch (BHB) stars, using MegaCam $g$- and $r$-band photometry. The third test targets RGB and BHB stars, and is based on \textit{Gaia} DR3 $G$- and $G_{RP}$-band photometry. Additionally, proper motion constraints in RA and Dec from \textit{Gaia} DR3 are incorporated in the second and third tests to further refine cluster membership. Finally, the fourth test uses DECam $g$- and $i$-band data, also targeting MS, RGB, and BHB stars. A summary of the characteristics of each test is presented in Table \ref{tab:tests_characteristics}.

Independently of these tests, we constructed LFs using MegaCam and DECam $g$-band photometry. Incorporating DECam data allows us to probe a larger spatial area than MegaCam, enabling the use of greater radial distances to establish the cluster's maximum detectable extent. 

\begin{table*}
\centering
\caption{Stars selected in each test, their photometry source, whether it includes proper motion constraints from \textit{Gaia} DR3, the covering area, and the number of stars selected.}
\label{tab:tests_characteristics}
\resizebox{0.86\textwidth}{!}{
\begin{tabular}{cccccc}
\hline
\hline
Test & Photometry & CMD stars & Proper motion data & Covering area & $\textup{N}_{\textup{stars}}$\\ 
& & & & ($\textup{arcmin}^{\textup{2}}$) & \\ \hline
1 & MegaCam & MS, RGB, BHB & - & $24\times24$ & $26179$\\
2 & MegaCam & RGB, BHB & \textit{Gaia} DR3 & $\pi \thinspace 15^{2}$ & $672$\\
3 & \textit{Gaia} DR3 & RGB, BHB & \textit{Gaia} DR3 & $\pi\thinspace60^{2}$ & $792$\\ 
4 & DECam & MS, RGB, BHB & - & $\pi\thinspace35^{2}$ & $5029$\\
\hline
\end{tabular}
}
\end{table*}

\subsection{Star membership identification}\label{subsec:members}

\begin{figure}[h]
    \centering
    \includegraphics[width=\columnwidth]{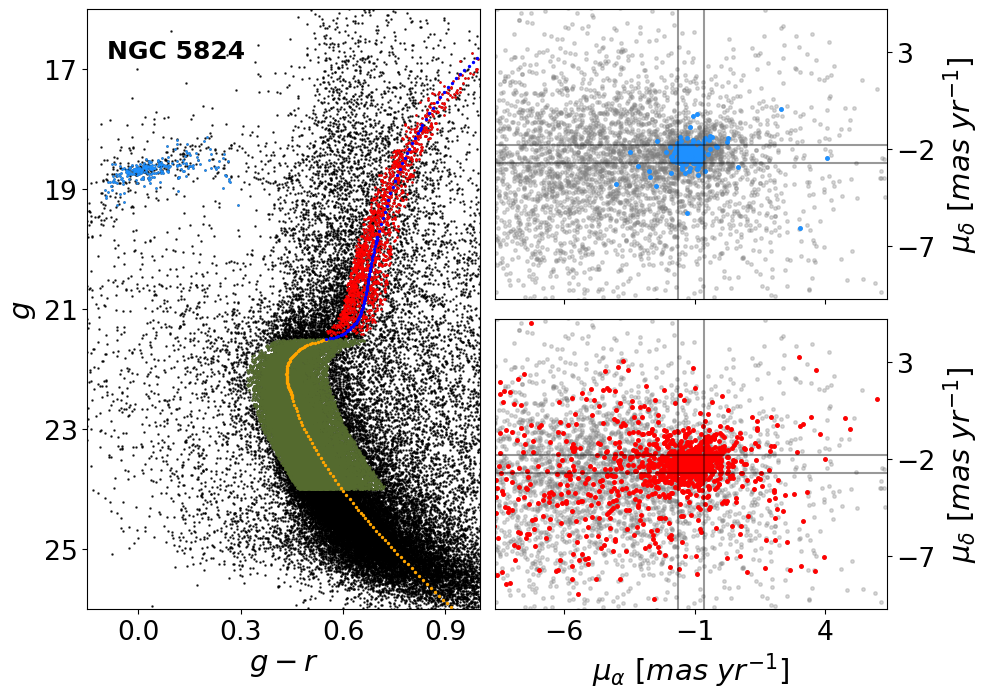}
    \caption{Left: CMD of NGC\,5824 with Clay photometry. Black dots correspond to the full \textit{sharp}-cleaned data. Light blue dots are BHB candidates. Red dots are RGB candidates. Blue and yellow dots correspond to the isochrone. Right: NGC\,5824 proper motion (PM) diagrams. Gray dots correspond to the full data with PM measurements. Light blue dots (upper panel) are BHB candidates. Red dots (lower panel) are RGB candidates.}
    \label{fig:ngc5824_cmd_pm}
\end{figure}

For the first test, we used the photometric catalog of Clay-MegaCam from \citet{Munoz_2018b} and constructed $g$ versus $g-r$ color-magnitude diagrams (CMDs). We used these CMDs to select MS and RGB stars with the highest probability of belonging to NGC\,5824. For the MS selection, we considered stars with $21.5 \leq g \leq 24$. This upper magnitude limit was chosen because, first, at this magnitude, the completeness level is $\approx 90\%$ (see Fig. 2 from \citealt{Munoz_2010} for a reference), and second, it minimizes the contamination from unresolved galaxies, which begin to dominate beyond $g=22$, while maintaining low photometric errors (see Fig. \ref{fig:photometric_errors}). To refine the MS selection, we incorporated a dynamic maximum Euclidean distance in the CMD space, $d_{dyn}$, from the best-fit isochrone to the MS, that       is, a distance that increases to fainter magnitudes, which follows the equation
\begin{equation}\label{eq:dyn_distance_MS}
    d_{dyn} = d_0 + k\ |g-g_{min}|
,\end{equation}
where $d_0$ is the base minimum distance, measured in mag, $k$ is a dimensionless scale factor, and $g$ is the magnitude value in the $g$ band. After some experimentation, we adopted the following values: $d_0 = 0.11$, $k = 0.0095$, and $g_{min}=21.5$. The values chosen here, and for subsequent star selections, best trace the observed features of the sequences. 

The isochrones used in this test and hereafter were obtained from the Dartmouth Stellar Evolution Database\footnote{\url{http://stellar.dartmouth.edu/models/isolf_new.html}} \citep{Dartmouth_2008}. Additionally, note that the adopted isochrone parameters are used only to facilitate the GC's star selection, rather than to determine any physical properties of the cluster. Particularly, for this test we used the CFHT-MegaCam ugriz photometric system, with ${\rm age = 13 \thinspace Gyr}$, ${\rm [Fe/H] = -1.91}$, $\rm {d_{\sun} = 42\,kpc}$ and ${\rm E(g-r) = 0.25}$.

We considered RGB stars as those with $13.6 \leq g \leq 21.5$ and selected them with the same dynamical distance criterion from Eq. \ref{eq:dyn_distance_MS}, but with different parameters: $d_0 = 0.04$, $k = 0.0027$, and $g_{min}=13.6$. We used the same isochrone, but with a different extinction value of ${\rm E(g-r) = 0.24}$. We also included stars in the BHB region of the CMD, a region typically with very low MW background contamination. BHB stars were selected by defining a specific CMD region that encloses the BHB locus, with boundaries in color $g-r$ and in magnitude $g$ sufficiently narrow to minimize contamination. Specifically, stars with $18.38 \leq g \leq 18.9$ and $-0.09 \leq g-r \leq0.18$ were classified as BHB stars.

\begin{figure*}
    \centering

    \begin{subfigure}[b]{\columnwidth}
        \centering
        \includegraphics[width=\textwidth]{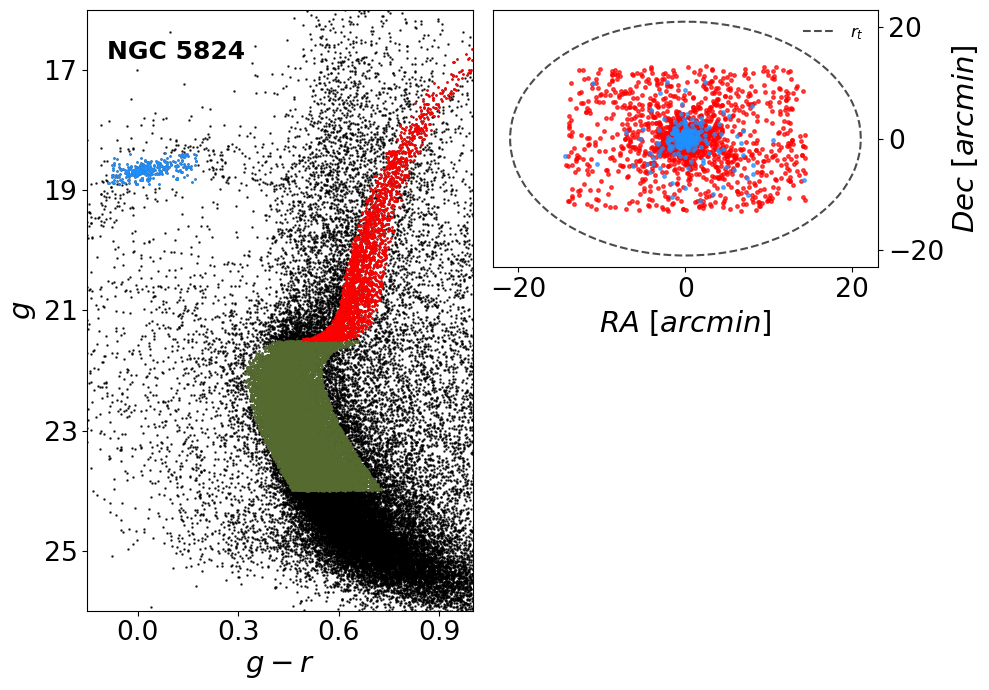}
    \end{subfigure}
    \hfill
    \begin{subfigure}[b]{\columnwidth}
        \centering
        \includegraphics[width=\textwidth]{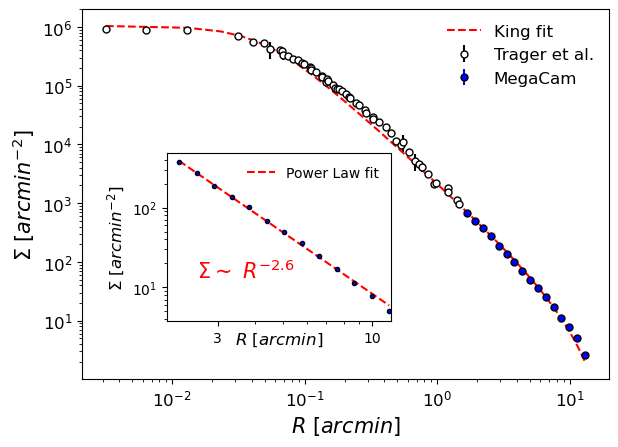}
    \end{subfigure}

    \vspace{0.5em}
    
    \begin{subfigure}[b]{\columnwidth}
        \centering
        \includegraphics[width=\textwidth]{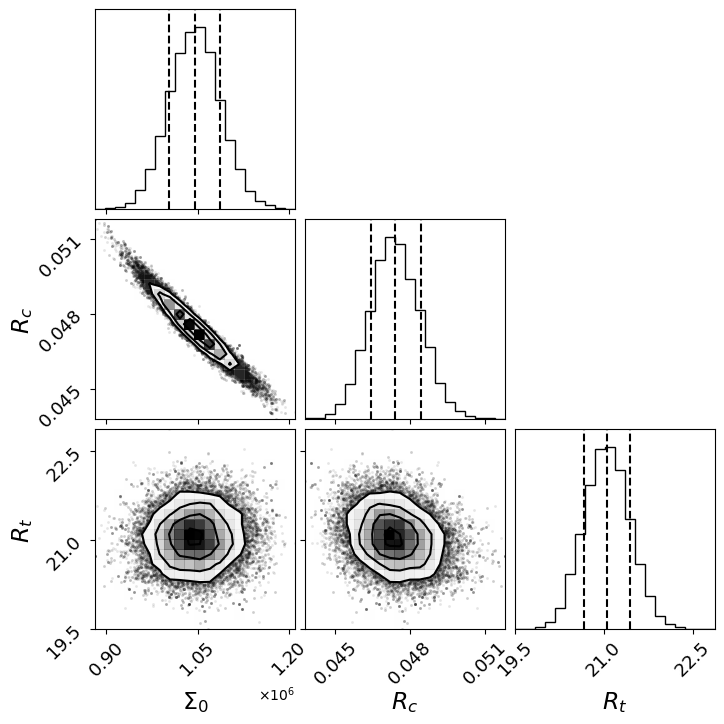}
    \end{subfigure}
    \hfill
    \begin{subfigure}[b]{\columnwidth}
        \centering
        \includegraphics[width=\textwidth]{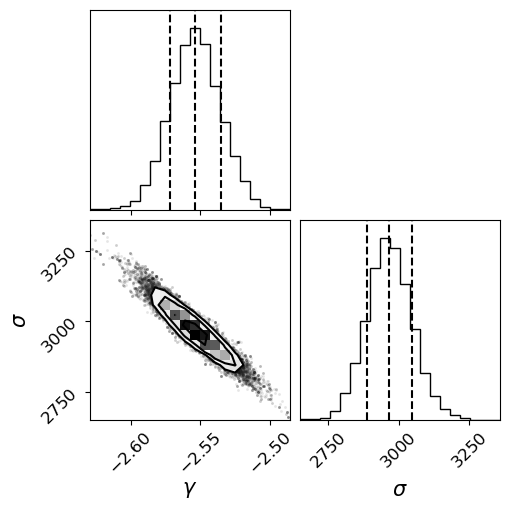}
    \end{subfigure}
    
    \caption{NGC\,5824 results for Test 1.Top left: CMD and a star-count map of the cluster.
 Green, red, and blue dots correspond to the MS, RGB, and BHB star candidates, respectively, and the gray circle represents the best-fit $r_t$ parameter for this test. Note that MS candidates are omitted from the star-count map because they occupy the entire MegaCam field. Top right: Observed surface number density profile, constructed using MS, RGB, and BHB stars, along with data from \cite{Trager_1995} for the cluster's innermost region. Bottom left and bottom right: Corner plots obtained for the King profile and power-law fits, respectively. The dashed black lines correspond to percentiles $0.16$, $0.5$, and $0.83$, in that order, from left to right. $R_c$ and $R_t$ are expressed in arcminutes. $\Sigma_0$ is expressed in $\rm{arcmin}^{-2}$. $\gamma$ is dimensionless, and $\sigma$ is a normalization factor of the power-law model.}
    \label{fig:test1_results}
\end{figure*}

Thus, for this first test, we considered stars from the MS, RGB and BHB. The left panel of Fig. \ref{fig:ngc5824_cmd_pm} shows the cluster's CMD and the stars selected for this test. To evaluate the sensitivity of the results to the selection criteria, we applied different constraints on the \textit{sharp} parameter and MS star selection.

First, we applied a more restrictive \textit{sharp} filter while maintaining the original MS selection shown in Fig. \ref{fig:ngc5824_cmd_pm}. Then, we compared two extreme cases, one minimizing and the other increasing photometric errors, through a variation in the depth of the MS selection between $g=23$ and $g=25$ while keeping the original \textit{sharp} filtering illustrated in Fig. \ref{fig:Sharp}. Finally, the original MS depth ($g=24$) and \textit{sharp} filter (Fig. \ref{fig:Sharp}) were fixed but with adjusted MS width, testing both narrower (using $d_0 = 0.07$) and wider (using $d_0 = 0.15$) selections than that shown in Fig. \ref{fig:ngc5824_cmd_pm}.

For the second test, we combined data from the Clay catalog with data from \textit{Gaia} DR3 \citep{gaia_2023} to aid the identification of distant cluster stars using proper motion and CMD properties. We first selected stars from the Clay-MegaCam catalog with $g < 21.5$ that have \textit{Gaia} DR3 proper motion measurements in both right ascension ($\mu_{\alpha}^{*}$)\footnote{The asterisk in $\mu_{\alpha}^{*}$ denotes the proper motion in right ascension corrected by declination, i.e., $\mu_{\alpha}^{*}=\mu_{\alpha}cos(\delta)$. This notation is omitted in the figures for simplicity.} and declination ($\mu_{\delta}$). This selection was made using the TOPCAT\footnote{\url{https://www.star.bris.ac.uk/~mbt/topcat/}} software, where we matched stars from Clay-Megacam with the ones in the \textit{Gaia} DR3 catalog by position (right ascension and declination), allowing a maximum error of $1''$ between the positions. In contrast to the first test, we did not include the MS stars since the \textit{Gaia} DR3 limiting magnitude is $G_{lim} = 20.7$ mag \citep{gaia_edr3}, which is brighter than for Clay-MegaCam data.

From this subset, the first filter used to select NGC\,5824 star candidates was based on the CMD positions of the stars. We considered the stars from the BHB and RGB of the CMD as shown in the left panel of Fig. \ref{fig:ngc5824_cmd_pm}. For this selection, we adopted the same methodology as for Test 1.

We subsequently used the proper motion information to further refine the filter used to select cluster stars. Since the BHB region of the CMD is relatively clean of MW stars, we used this group of stars to provide a reliable representation of NGC\,5824's proper motion. Figure \ref{fig:ngc5824_cmd_pm} (upper right panel) shows a well-defined area in proper motion space, further confirming their status as representative stars. 

The lower right panel of Fig. \ref{fig:ngc5824_cmd_pm} shows the proper motion of RGB stars (marked in red in the CMD). The region in proper motion space is similar but with a wider distribution, likely due to the presence of MW foreground stars and the fact that the RGB extends to fainter magnitudes than the BHB and therefore the proper motion errors for these stars are larger. To refine the selection of NGC\,5824 star candidates, we defined a proper motion box (using the BHB star candidates region as reference) to be: $\mu_{\alpha}^{*}$ $\in$ [-1.65, 0.65]\,$\rm{mas}$\,$\rm{yr^{-1}}$ and $\mu_{\delta}$ $\in$ [-2.75, -1.8]\,$\rm{mas}$\,$\rm{yr^{-1}}$, which is in agreement with the systemic proper motion found by \cite{Yang_2022}. 

In the third test, we used both photometry and proper motions from the \textit{Gaia} DR3 catalog only, up to $60'$ from the cluster's center. The procedure to select GC stars is as follows: First we filtered the stars by proper motion within the same specified range detailed above. Then, we selected BHB and RGB stars according to their CMD location and their proximity to the best-fit isochrone, respectively. For this isochrone, we used the \textit{Gaia} DR2 Revised photometric system, with ${\rm age = 13 \thinspace Gyr}$, ${\rm [Fe/H] = -1.5}$, $\rm {d_{\sun} = 35\,kpc}$ and ${\rm E(g-r) = 0.11}$. To facilitate the RGB selection, we adopted the same dynamical distance criterion of Eq. \ref{eq:dyn_distance_MS} used in Test 1 and 2, using $d_0 = 0.07$, $k = 0.005$ and $g_{min}=14.8$. 

Finally, for the fourth test, we used DECam photometry from \cite{Kuzma_2018} covering up to $35'$ from the cluster's center, which corresponds to $2.5$ times its King tidal radius, according to the measurement by \citet{Munoz_2018b}. The star selection procedure is similar to Test 1. The isochrone used was in the DECam photometric system, with ${\rm age = 13 \thinspace Gyr}$, ${\rm [Fe/H] = -1.91}$, $\rm {d_{\sun} = 32.1\,kpc}$ and distinct extinction values for MS and RGB selections, ${\rm E(g-r) = 0.12}$ and ${\rm E(g-r) = 0.06}$, respectively. MS stars, with $21.05\leq g \leq 22.3$, and RGB stars, with $17 \leq g \leq 21.04$, were selected incorporating the dynamical distance criterion, using $d_0 = 0.09$, $k = 0.03$ and $g_{min}=21.05$ for MS stars, and $d_0 = 0.035$, $k = 0.002$ and $g_{min}=17$ for RGB stars.

For the case of NGC\,2419, used as a control GC, we followed a procedure similar to that used in the first test (Clay-MegaCam catalog). We constructed the NGC\,2419 CMD using the \textit{sharp}-filtered stars from Fig. \ref{fig:Sharp}. Then, we selected the stars located in the BHB and RGB, the latter with $18 \leq g \leq 23.38$, selected within a fixed distance of $0.05$\,mag from the best-fit isochrone, and stars in the MS, with $23.4 \leq g \leq 25$, within a dynamic distance $d_{dyn}$ from the isochrone as specified in Eq. \ref{eq:dyn_distance_MS} using $d_0 = 0.013$, $k = 0.005$ and $g_{min}=23.4$. The isochrone is also in the CFHT-MegaCam ugriz photometric system, as for Tests 1 and 2, but with ${\rm age = 12 \thinspace Gyr}$ and ${\rm [Fe/H] = -1.5}$, $\rm {d_{\sun} = 98\,kpc}$ and ${\rm E(g-r) = 0.04}$. We extended the MS selection to fainter magnitudes than for NGC\,5824 because adopting a brighter limit would trace a limited portion of the sequence. Therefore, this choice is motivated by a trade-off between magnitude depth and minimizing photometric errors, which remain low at magnitude $g=25$ as illustrated in lower panels of Fig. \ref{fig:photometric_errors}.

\subsection{Density profiles}\label{subsec:density_profiles}

To construct the radial density profiles, we spatially divided the previously selected GC stars into several rings, centered on the GC coordinates, using a geometric progression, and computed the surface number density in each ring. We subtracted the average background surface density contribution $\Sigma_b$ from each surface density value of the GC. This background is determined in different ways depending on the area covered by the selected stars.

In the case of NGC\,5824, for the first and second tests, we determined the background by averaging the surface density measured in rectangular areas at the corners of the cluster's field. The idea is to use an area as far as possible from the cluster's center, in order to include as few cluster stars as possible (assuming they extend up to the edge of our covered area). For the third and fourth tests, considering the area covered extends out to a radius of $60'$ and $35'$, respectively, we calculated the background using the density of the maximum concentric circular annuli. For NGC\,2419, we used the same method as for the first and second tests for NGC\,5824.

An important consideration is that the photometry used in this study suffers from overcrowding effects at the innermost regions of the clusters. For this reason, we used the surface brightness profiles of NGC\,5824 and NGC\,2419 from \citet[hereafter Trager]{Trager_1995} to complement our profiles at small radii. To ensure consistency in the units, we converted this brightness density profile to a number density profile, i.e., converting measurements from flux units to star-count units. A cubic spline interpolation between Trager data and Clay/CFHT-MegaCam, \textit{Gaia} DR3, or DECam data, depending on the case, was conducted to facilitate the conversion. With the interpolated surface number density values, the following conversion from flux to star-counts was used \citep{Carballo-Bello_2011}:
\begin{equation}\label{eq:flux_number_transform}
    \rho = c - 0.4 \mu
,\end{equation}
where $\rho$ represents the interpolated Trager number surface density values, $\mu$ is the associated Trager flux surface density value, and $c$ is a constant. 

Instead of providing statistical uncertainties, Trager subjectively assigned to each data point $i$ a weight $w_i \in [0,1]$, reflecting the quality of the measurement. These weights, however, are not clearly related to conventional statistical error bars. To estimate realistic uncertainties for the surface brightness values $\delta\mu_{i}$, where $i$ represents each Trager data point, the methodology proposed by \cite{McLaughlin_2005} was implemented, which assumes that the errors are inversely proportional to the assigned weights, i.e., $\delta\mu_{i} \propto 1/w_i$. This approach provides a statistically and physically consistent calibration of the uncertainties for each cluster. The proportionality constant for the case of NGC\,5824 is $0.106$. Therefore, the individual uncertainties for each Trager density data point are then computed as $\delta\mu_{i} = 0.106/w_i$.

Finally, we fitted a King profile to the number density profiles for each test of NGC\,5824 and NGC\,2419. This King profile follows the following equation:
\begin{equation}\label{eq:King}
    \Sigma_{King}(R) = \Sigma_0\left[\left( 1 + \frac{R^2}{R_{c}^{2}}\right)^{-\frac{1}{2}} - \left( 1 + \frac{R_{t}^{2}}{R_{c}^{2}}\right)^{-\frac{1}{2}} \right]
,\end{equation}
where $\Sigma_0$ is the central surface density of the cluster, $R_c$ is its core radius and $R_t$ is its King tidal radius.  

For the parameter estimation procedure, we used the \textit{emcee} library minimizing a $\chi^2$ and exploring the parameter space following the rules of Markov chain Monte Carlo (MCMC). In the end, we obtained a corner plot that shows the distribution of each fitted parameter and the relations between them. These plots are shown in Figs. \ref{fig:test1_results}, \ref{fig:test2_results}, and \ref{fig:test3_results} for Tests 1, 2, and 3, respectively.

\begin{table*}
\centering
\caption{Background density and best-fitting parameters.}
\label{tab:best_fit_params}
\resizebox{\textwidth}{!}
{
\begin{tabular}{ccccccccc}
\hline\hline
Test & Background density & \multicolumn{5}{c}{Fitted King radii parameters} & \multicolumn{2}{c}{Fitted power-law index} \\
 & $\Sigma_b$ & $R_c$ & $R_t$ & $R_c$ & $R_t$ & $\chi^2_{King}$ & \multicolumn{1}{c}{$\gamma$} & $\chi^2_{pw}$ \\
 & ($\textup{arcmin}^{-2}$) & (arcmin) & (arcmin) & (pc) & (pc) & & &  \\ \hline
 \multicolumn{9}{c}{NGC\,5824}\\
 \hline
1 & $1.60\pm0.22$ & $0.047^{+0.001}_{-0.001}$ & $21.04^{+0.39}_{-0.38} $ & $0.44^{+0.01}_{-0.01}$ & $196.5^{+3.6}_{-3.5}$ & $282.7$ & $-2.55^{+0.02}_{-0.02}$ & $23.8$ \\
2 & -- & $0.055^{+0.001}_{-0.001}$ & $16.725^{+1.004}_{-0.967}$ & $0.51^{+0.01}_{-0.01}$ & $156.2^{+9.4}_{-9.0}$ & $90.5$ & $-2.49^{+0.11}_{-0.11}$ & $2.9$\\
3 & $0.016\pm0.002$ & $0.052^{+0.001}_{-0.001}$ & $28.5^{+0.8}_{-0.8}$ & $0.49^{+0.01}_{-0.01}$ & $266.1^{+4.7}_{-4.7}$ & $188.2$ & $-2.48^{+0.08}_{-0.08}$ & $11.4$\\
4 & $0.47\pm0.02$ & $0.048^{+0.001}_{-0.001}$ & $20.95^{+0.32}_{-0.31}$ & $0.45^{+0.01}_{-0.01}$ & $196^{+3}_{-3}$ & $413.7$ & $-2.79^{+0.12}_{-0.14}$ & $0.4$\\
\hline
\multicolumn{9}{c}{NGC\,2419}\\
\hline
 1 & $0.72\pm0.03$ & $0.253^{+0.004}_{-0.004}$ & $10.0^{+0.1}_{-0.1}$ & $6.1^{+0.1}_{-0.1}$ & $240.3^{+2.4}_{-2.4}$ & $213.1$ & $-4.47^{+0.09}_{-0.09}$ & $36.7$ \\
\hline
\end{tabular}
}
\end{table*}

To characterize the outskirts of the clusters region, we fitted a power-law profile only to the points corresponding to the outer region of the profiles for the four different tests. This profile follows the equation
\begin{equation}\label{eq:pw}
    \Sigma_{pl}(R) = \sigma R^{\gamma}
,\end{equation}
where $\gamma$ is the power-law index that represents the slope of the profile, and $\sigma$ is a normalization factor. Additionally, the $\chi^2$ values for both the King ($\chi^2_{King}$) and power-law ($\chi^2_{pw}$) models were computed for each test to evaluate which model better fits the data.

\subsection{g-band luminosity function}\label{subsec:LFs}

To estimate the approximate extent of NGC\,5824 and to assess the presence of a diffuse stellar envelope, we analyzed cumulative $g$-band LFs, tracing the turn-off point and MS of the cluster across different annuli. These LFs were compared with the normalized background cumulative LF to assess the radial distance from the cluster center at which we can reliably detect cluster members above the background level. For this analysis, we separately tested MegaCam and DECam photometry.

\begin{figure}[h]
    \centering
    \includegraphics[width=\columnwidth]{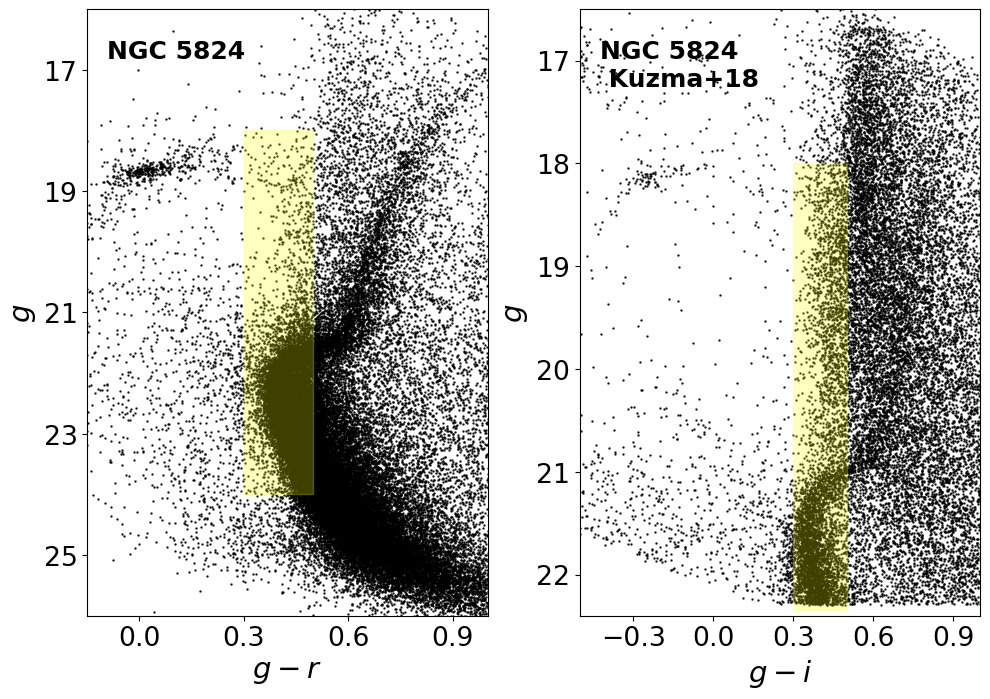}
    \caption{CMDs of NGC\,5824 based on  MegaCam photometry (left) and DECam photometry (right). In the MegaCam CMD, black dots correspond to the full \textit{sharp}-cleaned data, while in the DECam CMD, black dots correspond to stars within $1'\leq R \leq 35'$, where R is the angular distance from the cluster center. The yellow shaded regions indicate the color and magnitude selection used to construct the cumulative LFs.}
    \label{fig:ngc5824_LF_lims}
\end{figure}

The selection of stars used to construct the cumulative LFs follows a color range in $(g-r)$ for MegaCam and $(g-i)$ for DECam, and a $g$ magnitude filter as shown in Fig. \ref{fig:ngc5824_LF_lims}. Specifically, MegaCam stars follow $18\,\leq g \leq\,24$ and $0.3\, \leq g-r \leq\,0.5$, while DECam stars follow $18\,\leq g \leq\,22.34$ and $0.3\, \leq g-i \leq\,0.5$. The chosen magnitude range is defined to ensure that we are accurately tracing the background, the turn-off and the MS of the cluster, while the chosen color range maximizes the contrast between cluster and background stars.

The MegaCam and DECam backgrounds were defined consistently with the approach used in the density profile section: for MegaCam, four squares located in the corners of the Clay-MegaCam field of view were used, while for DECam, the maximum concentric circular annulus was employed. Background stars were further divided into four quadrants. We then constructed and characterized the LFs of both the total background and its individual quadrants.

\begin{figure}[h]
    \centering
    \includegraphics[width=\columnwidth]{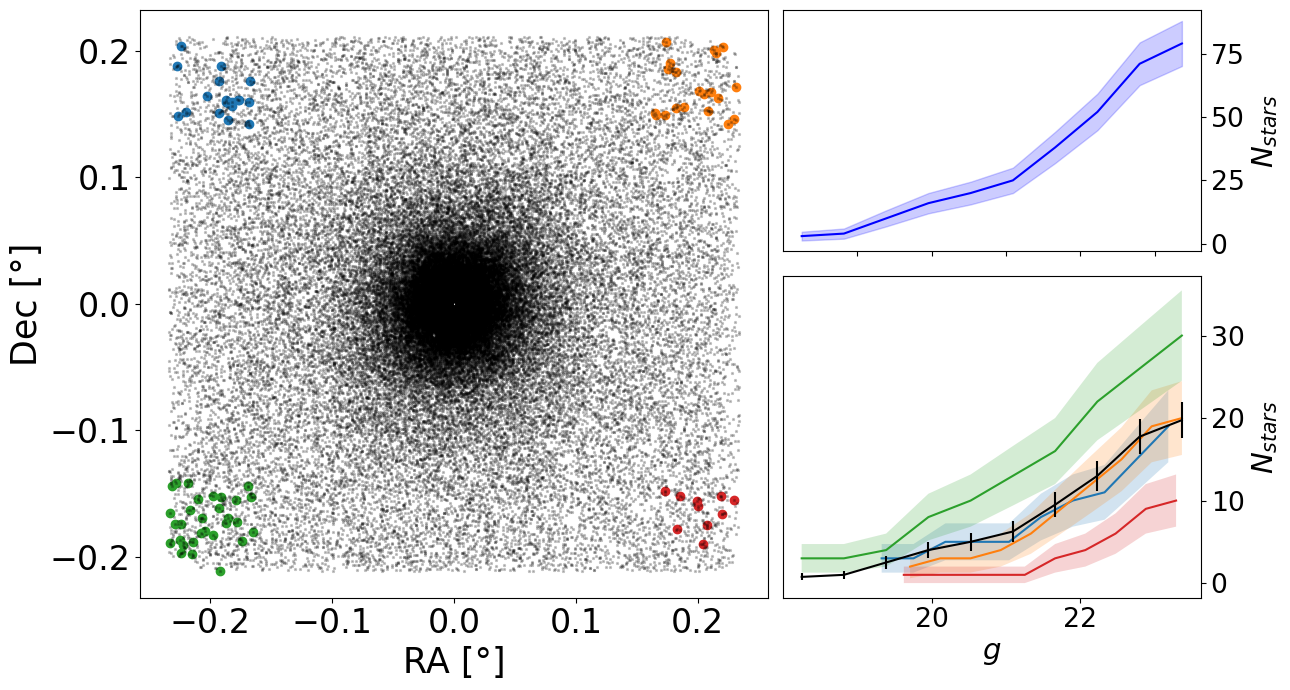}

    \vspace{0.5em}

    \includegraphics[width=\columnwidth]{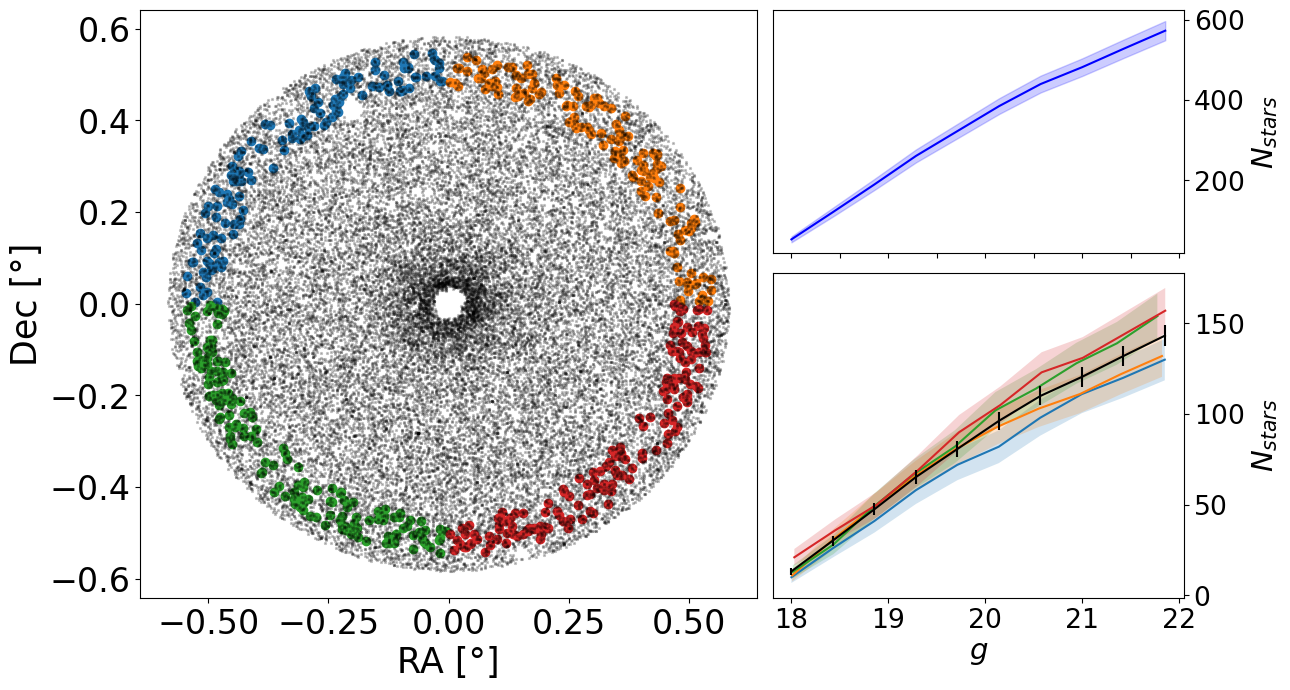}
    
    \caption{Background LFs of NGC\,5824 based on MegaCam (upper panel) and \cite{Kuzma_2018} DECam data (lower panel). Each panel contains a star-count map (left subpanel), the cumulative LF of the entire background (top right), and the cumulative LFs for each corner of the background stars, as indicated in the star-count map (bottom right). In the left subpanels, black dots correspond to the full \textit{sharp}-cleaned data (only for the MegaCam data). Colored dots represent the color filtered stars used as background. In the right subpanels, the solid black line corresponds to the total background divided by four. The shaded areas in both panels and the black error bars in the bottom panels represent the corresponding Poisson errors, $\sqrt{N_{stars}}$.}
    \label{fig:LFs_background}
\end{figure}

Using the selected cluster stars, we constructed the cumulative LFs for each annulus, each with a size of $3'$ for the MegaCam case (see Fig. \ref{fig:LFs_ngc5824}), and $5'$ for the DECam case (see Fig. \ref{fig:LFs_ngc5824_Kuzma}). To compare the cluster LF at different annuli with the background LF, the background LF was normalized at $g=20$, as there is no MS at this magnitude in either MegaCam or DECam photometry. We defined a contrast parameter ($C$) to compare the cluster and background cumulative LFs, defined as the ratio between the maximum LF value of the cluster ($N_{c}^{max}$) and the background ($N_{b}^{max}$):
\begin{equation}
    C = \frac{N_{c}^{max}}{N_{b}^{max}}
.\end{equation}

To assess the symmetry of the cluster, we analyzed the LF of the outermost ring that remained distinguishable from the background (see Fig. \ref{fig:LFs_quarter_ngc5824} for the MegaCam case and Fig. \ref{fig:LFs_quarter_ngc5824_Kuzma} for DECam case). This ring was divided into four quadrants: the upper-right, upper-left, lower-right, and lower-left quadrants. As a result of this division, the number of cluster stars decreased; therefore, we adopted a larger number of bins for constructing these LFs. To maintain consistency, the background LF was also constructed with the same number of bins. It is important to note that the background stars were not divided into quadrants, instead they represent the same overall background.

\section{Results}\label{results}
The star selection and density profile results for NGC\,5824 are presented in Figs. \ref{fig:test1_results}, \ref{fig:test2_results}, \ref{fig:test3_results}, and \ref{fig:test4_results} for Tests 1 through 4, respectively, and in Fig. \ref{fig:ngc2419_results} for NGC\,2419. For each test, the best-fitting parameters reported are the King limiting radius, $R_t$, and the slope $\gamma$. For NGC\,5824, $R_t \sim 21'$ for Test 1, $R_t \sim 17'$ for Test 2, $R_t \sim 29'$ for Test 3, and $R_t \sim 21'$ for Test 4. The power-law index remains consistent across all tests, with  an average $\gamma = - 2.5\pm0.1$. In the case of NGC\,2419, $R_t \sim 10'$, and the slope is steeper, with $\gamma \sim - 4.5$. Table \ref{tab:best_fit_params} summarizes the best-fit parameters for all tests. For both clusters, the $\chi^2_{pw}$ values are consistently lower than $\chi^2_{King}$, meaning that a power-law profile better describes the outer region of the cluster.

\begin{figure}[t]
    \centering
    \includegraphics[width=\columnwidth]{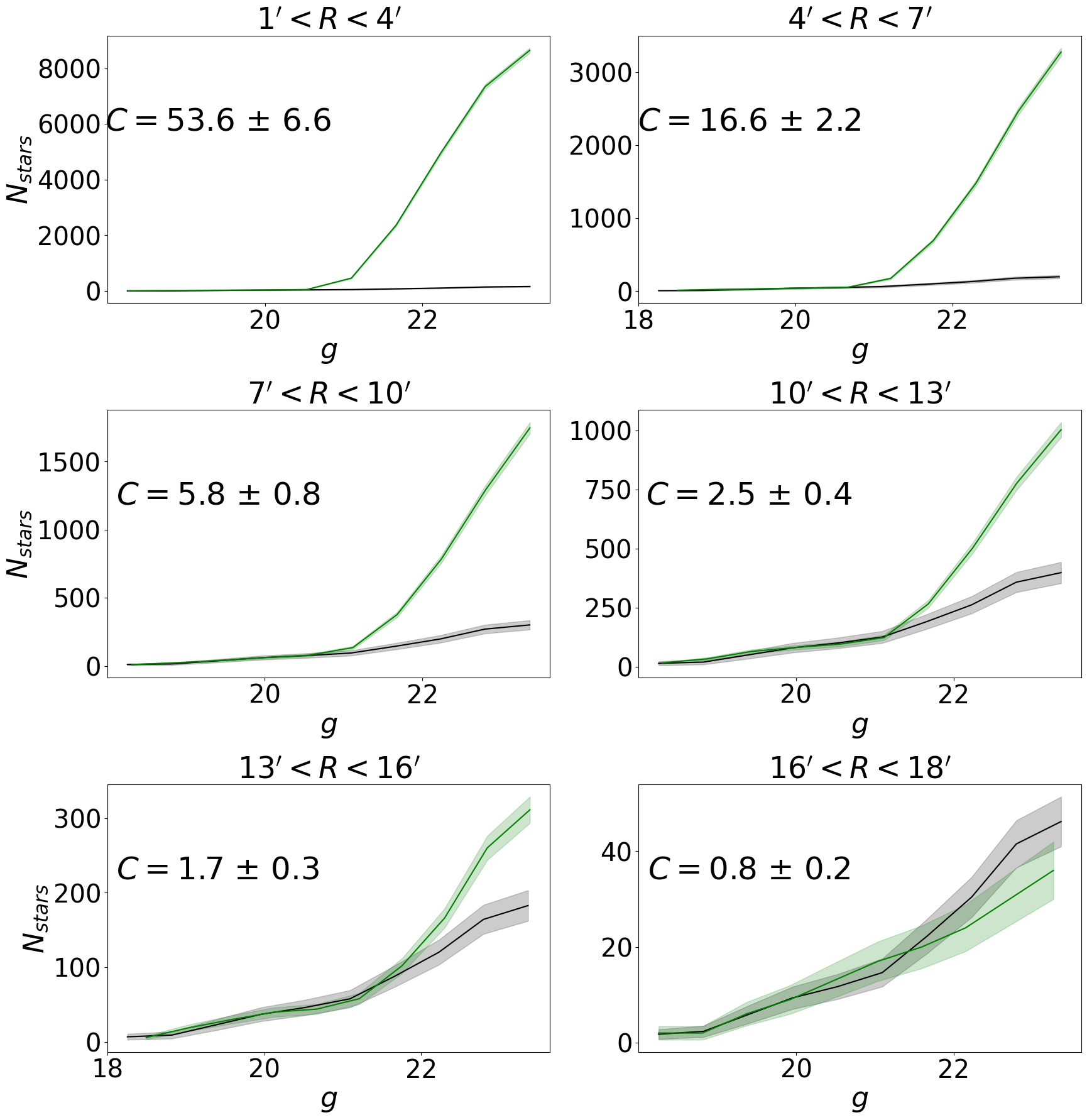}
    \caption{MegaCam cumulative LFs of NGC\,5824 for different rings. The green line and the black line represent the cluster and the background, respectively. The shaded area corresponds to the Poisson error bars ($\sqrt{N_{stars}}$). $C$ is the contrast parameter.}
    \label{fig:LFs_ngc5824}
\end{figure}

\begin{figure}[t]
    \centering
    \includegraphics[width=\columnwidth]{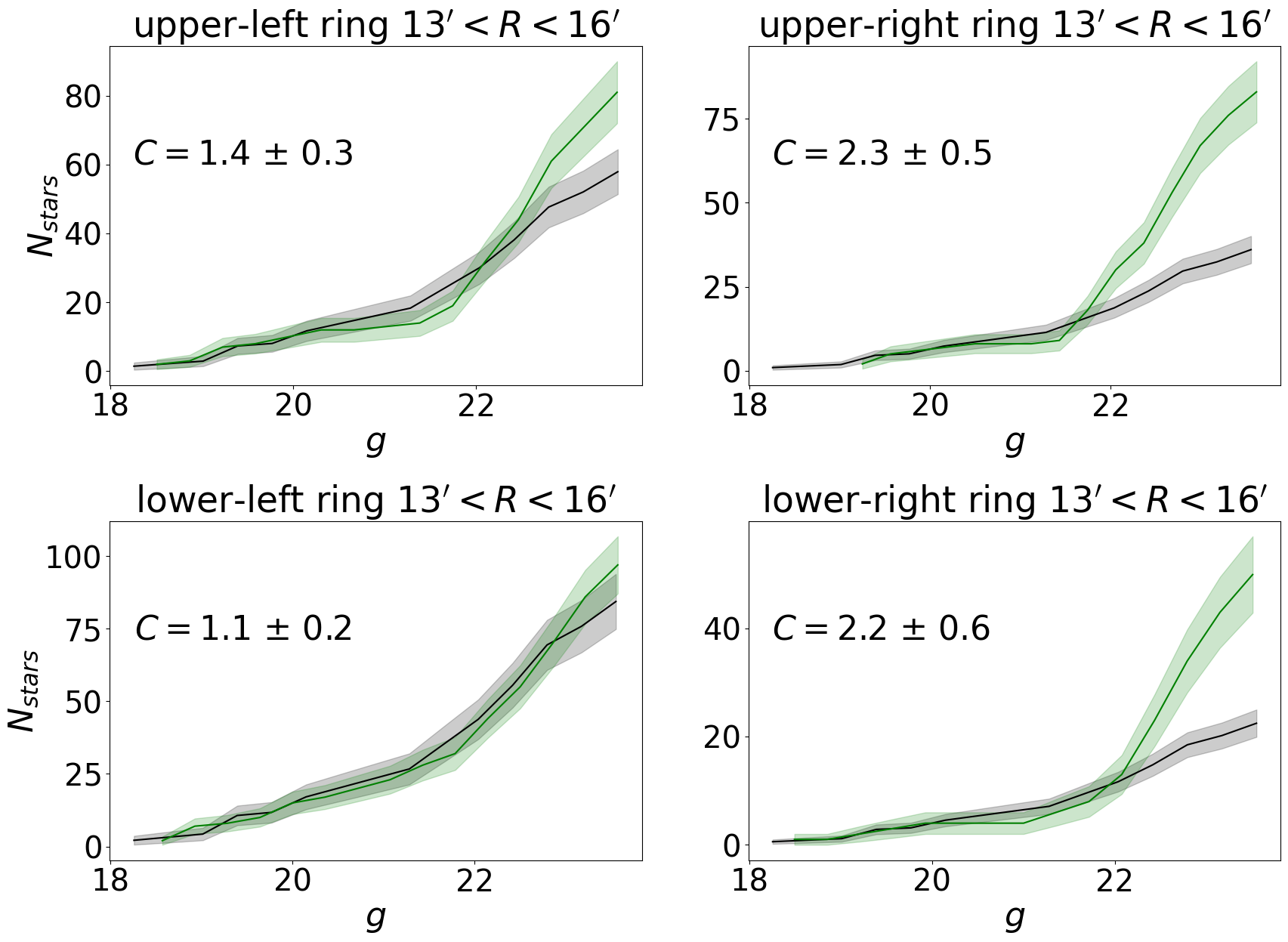}
    \caption{MegaCam cumulative LFs for quadrants of the ring $13' < R < 16'$. The colors have the same meanings as in Fig. \ref{fig:LFs_ngc5824}.}
\label{fig:LFs_quarter_ngc5824}
\end{figure}

The evaluation of the sensitivity for Test 1 showed that, regardless of the star selection criteria, the King and power-law profile fits remain stable, with only minor variations that do not affect the main conclusions. Specifically, the power-law index ($\gamma$) varies between $-2.7$ and $-2.4$, consistent with the statement by \citet{Penarrubia_2017}. The consistency of power-law fits across all tests further supports the robustness of these results, as they are derived from independent datasets.

Notably, in Test 2, NGC,5824 stars are detected out to a radius of $147$,pc ($\sim 15'.7$ in the MegaCam field). This extended stellar distribution is clearly visible in the star-count map shown in the upper-left panel of Fig. \ref{fig:test2_results}.

Finally, with respect to the NGC\,5824 $g$-band LFs, the MegaCam LFs for rings $1'<R<4'$ to $16'<R<18$ are shown in Fig. \ref{fig:LFs_ngc5824}, with LFs of the quadrants for the ring $13'<R<16'$ displayed in Fig. \ref{fig:LFs_quarter_ngc5824}. Similarly, the DECam LFs for rings $5'<R<10'$ to $30'<R<35$ are shown in Fig. \ref{fig:LFs_ngc5824_Kuzma}, with LFs of the quadrants of the ring $15'<R<20'$ presented in Fig. \ref{fig:LFs_quarter_ngc5824_Kuzma}.

\section{Analysis and discussion} \label{analysis}

As seen in multiple studies \citep[e.g.,][]{McConnachie_2012, Munoz_2018b}, GCs typically do not extend beyond $\sim100$\,pc, although exceptions have been observed. For example, NGC\,1851 has an extended stellar envelope of $\sim240$\,pc \citep[][]{Olszewski_2009, Kuzma_2018, Zhang_2022}. Using the MegaCam data, we find that the LF analysis of NGC\,5824 detects MS stars out to the edge of the field, at $\sim 16'$ from its center (equivalent to $\sim 150$\,pc), as illustrated in Fig. \ref{fig:LFs_ngc5824}. In Fig. \ref{fig:LFs_quarter_ngc5824} we use the MegaCam data to show a deeper analysis of the cluster's outermost region where the MS is still detectable, from which we can conclude that this region is tracing a symmetric structure. We note that the increase in the contrast parameter $C$ in the east panels of this outermost ring could be attributed to the spatial variation in the background shown in Fig. \ref{fig:LFs_background}. 

\begin{figure}[t]
    \centering
    \includegraphics[width=\columnwidth]{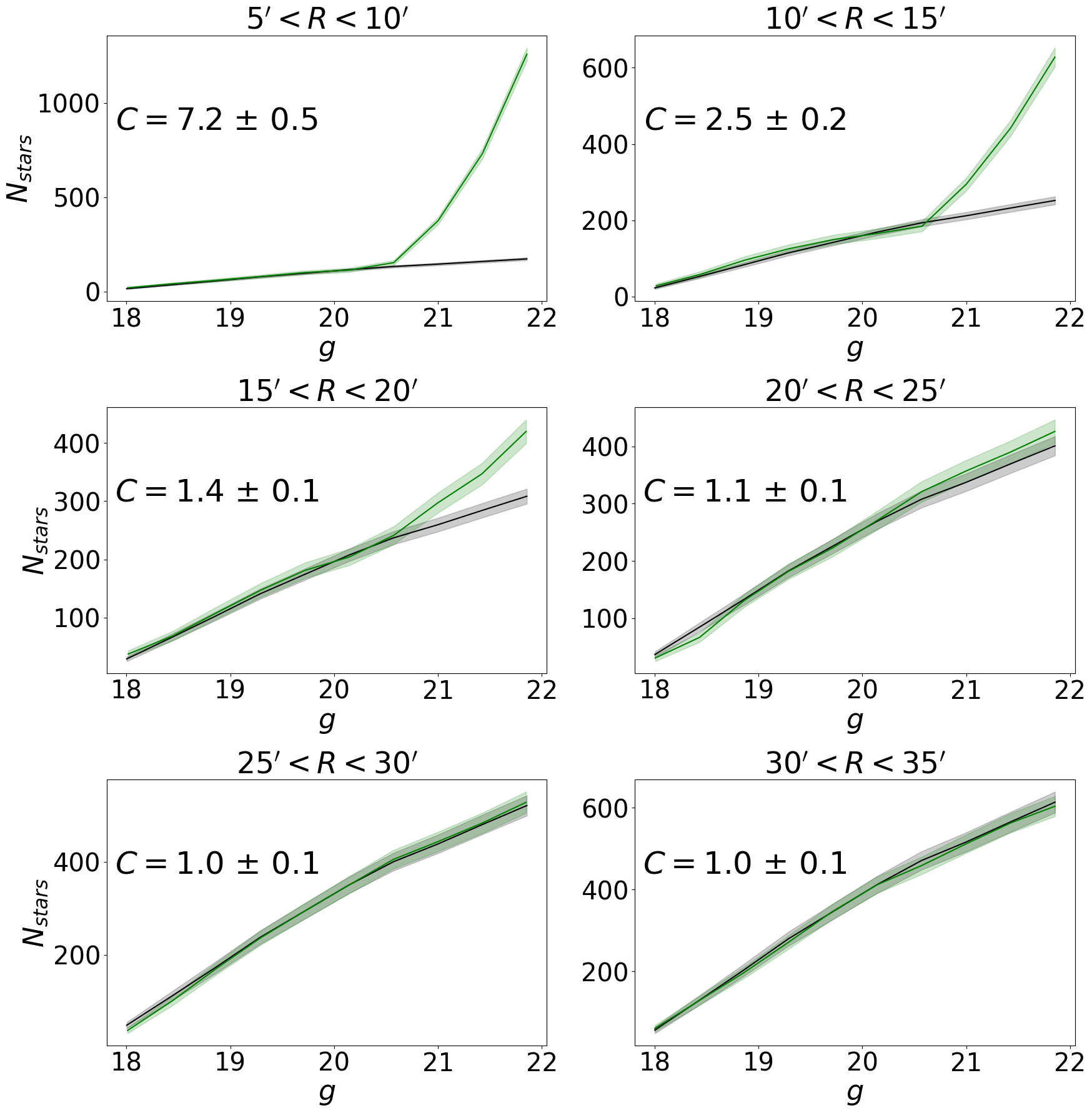}
    \caption{DECam cumulative LFs of NGC\,5824 for different rings. The colors have the same meanings as in Fig. \ref{fig:LFs_ngc5824}. The shaded area corresponds to the Poisson error bars ($\sqrt{N_{stars}}$). $C$ is the contrast parameter.}
\label{fig:LFs_ngc5824_Kuzma}
\end{figure}

\begin{figure}[t]
    \centering
    \includegraphics[width=\columnwidth]{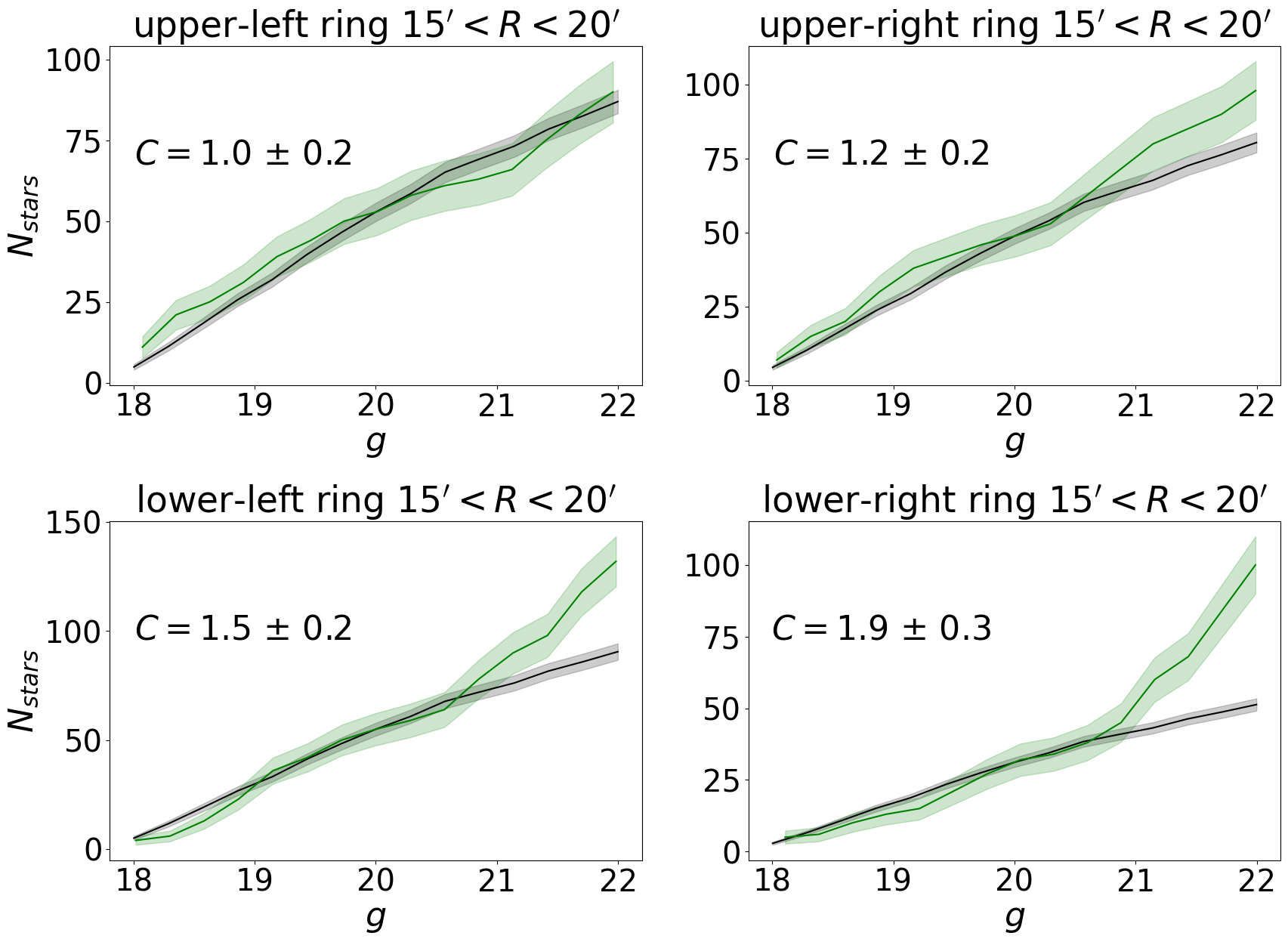}
    \caption{DECam cumulative LFs for quadrants of the ring $15' < R < 20'$. The colors have the same meanings as in Fig. \ref{fig:LFs_ngc5824}.}
\label{fig:LFs_quarter_ngc5824_Kuzma}
\end{figure}

On the other hand, \cite{Kuzma_2018} detected cluster stars out to $230$\,pc at a $3\sigma$ level of significance from 2D distribution imaging. Since MegaCam data have a limited radial coverage compared to \cite{Kuzma_2018}, we incorporated DECam data from their study into our analysis to test whether stars beyond MegaCam coverage can be detected using the LF methodology. According to Fig. \ref{fig:LFs_ngc5824_Kuzma}, cluster stars are clearly detected up to at least $20'$, while the ring between $20'$ and $25'$ suggests a marginal presence of stars beyond $20'$, as the cluster signal is slightly above the background in this region, consistent with the detection limit of $\sim24'$ reported by \cite{Kuzma_2018}. 

The King profile fits displayed in the top-right panels of Figs. \ref{fig:test1_results}, \ref{fig:test2_results}, and \ref{fig:test3_results} (corresponding to Test 1, 2, and 3, respectively) reveal an absence of a clear tidal cutoff. This absence may suggest that the MW has not yet truncated its stellar distribution. For comparison, we computed the cluster's Jacobi radius following Eq. 1 from \citep{Burkert_1997}, with the orbital parameters of the clusters taken from \cite{Bajkova_2021}. The Jacobi radius corresponds to the theoretical radius beyond which stars are unbound to the cluster. We assumed a Galactocentric distance of $25.6$\,kpc for NGC\,5824 and used the enclosed Galactic mass at this radius as described by Eq. 3 in \cite{Burkert_1997}, whose result is consistent with more recent MW mass profiles (see Fig. 11 from \citealt{Shen_2022}). 
We estimated the Jacobi radius to be $\sim 19'$, in line with the fitted King limiting radii. The fact that cluster stars are detected beyond these radii reinforces the notion that NGC\,5824 is unusually extended, presenting a ``halo" of stars beyond its expected tidal cutoff radius.

One possibility to explain these observations is to recall the \citet{Penarrubia_2017} model, where  the authors simulate GCs with varying amounts of DM, ranging from none, intermediate, and higher DM content. The evolution of the particles in the simulations is analyzed as they interact with the gravitational field of the system (stars+DM). The study reveals a clear trend in which the response of the density profile depends significantly on the amount of DM. For the zero DM case, the cluster develops steeper outer density profiles, while those with increasing DM content become more spatially extended (see Fig. 1 from \citealt{Penarrubia_2017}) and exhibit progressively shallower density profile slopes asymptotically approaching a value of $\sim -3$, without showing a clear tidal cutoff (see the upper panels of Fig. 2 from \citealt{Penarrubia_2017}).

Figures \ref{fig:test1_results}, \ref{fig:test2_results}, \ref{fig:test3_results}, and \ref{fig:test4_results} show that, at large distances, a power-law profile with a slope of $\gamma \sim -2.6$ is more appropriate than a King profile, which is consistent with the predictions of \citet{Penarrubia_2017}. In contrast, NGC\,2419 displays a much steeper slope of $\gamma \sim -4.5$ (see Fig. 
\ref{fig:ngc2419_results}). This steep profile is in line with the no DM scenario of \citet{Penarrubia_2017}. Moreover, the fact that power-law profiles systematically outperform King models in the outer regions, as in this case, is not new. \citet{Carballo-Bello_2011} already reported this behavior for several clusters, attributing it to the improved ability of modern surveys to detect stars in their low-density outskirts. Altogether, we argue that the findings of this work suggest that NGC\,5824 is a strong candidate to be embedded in a DM halo, which could account for its unusual extension and photometric properties. We note that there are other potential explanations for this observed extension. For example, if NGC\,5824 were approaching its pericenter, the Jacobi radius would shrink as the GC approaches the Galactic center, making stars that were initially bound now unbound. However, according to \cite{Bajkova_2021}, NGC\,5824's pericenter is $\sim 14.26$\,kpc, and currently the cluster is located at $25.6$\,kpc from the galactic center, not quite near the perigalacticon.

In a contrasting view, \cite{Yang_2022} proposed the presence of tidal debris in the vicinity of NGC\,5824. Notably, they claim the existence of a trailing tail observed to extend $\sim50$$\,deg$ from the cluster's center. Additionally, the authors propose that the Triangulum stream serves as a component of the leading tail associated with NGC\,5824. They compare the leading tail of a simulated stream of NGC\,5824 with Triangulum stream stars (see their Fig. 3). Although this comparison is tantalizing, the phase space distribution of the mock stars spans approximately $15\,deg$ in RA, which implies that the model is not precisely constrained. The authors argue that Triangulum stars are consistent with being associated with NGC\,5824 according to their metallicity and CMD positions. However, we consider that their match-filter map (see their Fig. 8), which is presented as the detection of the trailing tail of the cluster, shows partial detections of a structure that we cannot directly associate with the cluster as a distinctive structure indicative of a NGC\,5824 stream. More importantly, in their discussion, the authors acknowledge a significant inconsistency between Triangulum stream candidate stars and NGC\,5824 stars, i.e., according to their Fig. 10, most of the proposed members of the supposed leading tail stars lie at nearly twice the distance of the model prediction, corresponding to a difference of approximately 1.5 magnitudes. For all of the above, in our view the analysis presented in \cite{Yang_2022}, the existence and detection of tidal tails is plausible but not definitive to explain the extended stellar envelope of the cluster.

Although several mechanisms have been proposed to explain how stars in GCs reach large radii \citep[e.g.,][]{Weatherford_2023}, including evaporation, binary interactions, and supernova explosions, we did not attempt to identify the origin of such configurations. Instead, our analysis focused on the empirical fact that these stars exist and remain gravitationally bound to the cluster center at large distances, which we interpret as indicative of the presence of DM.

Based on the results presented above, we argue that NGC\,5824 is a good candidate to further investigate the presence of DM. Strong support for this hypothesis requires high-quality radial velocity profile data (see Figs. 2 and 3 from \citealt{Penarrubia_2017}), which are not yet available. Future studies addressing this problem could be key to resolving this question.

If the results of future observations were to confirm that DM is indeed present, one might ask whether NGC\,5824 should be classified instead as a dwarf galaxy, but the answer is not straightforward. Other factors, most notably NGC\,5824's chemically homogeneous stellar population, support the classification as a GC, leaving the door open to a reconsideration of what the fundamental boundaries between GCs and dwarf galaxies are.

\section{Conclusions} \label{conclusion}

In this study we analyzed the surface density profile of the GC NGC\,5824 using observations from the MegaCam imager at the Magellan Clay telescope in the $g$ and $r$ bands, with a limiting magnitude of $g_{lim} \simeq 25.6$. These data were complemented by the Trager catalog for the inner regions of the cluster, and by \textit{Gaia} DR3 $G$ and $G_{rp}$ bands for the outskirts. We also used \textit{Gaia} DR3 proper motions to constrain cluster stars. This combination of deep and homogeneous imaging with high-precision astrometry enhanced our ability to probe the extended stellar profile of NGC\,5824. Additionally, we incorporated DECam data from \cite{Kuzma_2018}, who report a cluster detection up to $230$\,pc, which is highly unusual for a GC.

Our primary goal was to assess the model predictions from \citet{Penarrubia_2017} regarding the outer slopes of GCs with or without DM content. We also examined the cumulative LF to assess the cluster's spatial symmetry and extent. As a control case, we also analyzed the outer halo GC NGC\,2419, for which we also had observations from the MegaCam imager at the CFHT.

 Previous studies of the GC NGC\,5824 indicate that the cluster presents multiple unusual characteristics, including an extended stellar distribution and the absence of an observed King tidal cutoff. Our results align with these findings, further affirming the unique nature of NGC\,5824 in this regard. From our results shown in Figs. \ref{fig:test1_results}, \ref{fig:test2_results}, \ref{fig:test3_results}, and \ref{fig:test4_results}, we find that in the outermost regions of NGC\,5824, a power-law profile is more appropriate than a King profile. Moreover, we estimate an average slope of $-2.6\pm0.1$ based on several different tests using three independent datasets, which is consistent with the predictions. Additionally, the cumulative LFs displayed in Figs. \ref{fig:LFs_ngc5824} and \ref{fig:LFs_quarter_ngc5824} for MegaCam data, as well as in Figs. \ref{fig:LFs_ngc5824_Kuzma} and \ref{fig:LFs_quarter_ngc5824_Kuzma} for DECam data from \cite{Kuzma_2018}, further corroborate the cluster's spatial symmetry and extended structure.

Based on these findings, we conclude that the density profile of NGC 5824 is unusual and corroborate the predictions made by \citet{Penarrubia_2017}. The absence of truncation in its density profile supports the idea that it still retains a DM component. However, kinematic data are necessary to confirm this. Future spectroscopic observations of NGC\,5824 stars in the outskirts will provide radial velocities, which will be critical for distinguishing whether the extended structure is due to DM or other dynamical effects.

\begin{acknowledgements}
P.B.D, B.M, R.R.M and V.S gratefully acknowledges support by the Agencia Nacional de Investigaci\'on y Desarrollo (ANID) Programa de Financiamiento Basal para Centros Cient\'ificos y Tecnol\'ogicos de Excelencia (BASAL) project FB210003. JAC-B acknowledges support from FONDECYT Regular N 1220083. PBK is supported by a UKRI Future Leaders Fellowship (MR/S018859/1).
      
This paper is based on observations obtained with the MegaPrime/MegaCam, a joint project of the Canada--France--Hawaii Telescope (CFHT) and CEA/DAPNIA, at CFHT, which is operated by the National Research Council (NRC) of Canada, the Institut National des Science de l'Univers of the Centre National de la Recherche Scientifique (CNRS) of France, and the University of Hawaii.

This project used data obtained with the Dark Energy Camera (DECam), which was constructed by the Dark Energy Survey (DES) collaborating institutions: Argonne National Lab, University of California Santa Cruz, University of Cambridge, Centro de Investigaciones Energeticas, Medioambientales y Tecnologicas-Madrid, University of Chicago, University College London, DES-Brazil consortium, University of Edinburgh, ETH-Zurich, University of Illinois at Urbana-Champaign, Institut de Ciencies de l'Espai, Institut de Fisica d'Altes Energies, Lawrence Berkeley National Lab, Ludwig-Maximilians Universitat, University of Michigan, National Optical Astronomy Observatory, University of Nottingham, Ohio State University, University of Pennsylvania, University of Portsmouth, SLAC National Lab, Stanford University, University of Sussex, and Texas A\&M University. Funding for DES, including DECam, has been provided by the U.S. Department of Energy, National Science Foundation, Ministry of Education and Science (Spain), Science and Technology Facilities Council (UK), Higher Education Funding Council (England), National Center for Supercomputing Applications, Kavli Institute for Cosmological Physics, Financiadora de Estudos e Projetos, Fundação Carlos Chagas Filho de Amparo a Pesquisa, Conselho Nacional de Desenvolvimento Científico e Tecnológico and the Ministério da Ciência e Tecnologia (Brazil), the German Research Foundation-sponsored cluster of excellence "Origin and Structure of the Universe" and the DES collaborating institutions.
\end{acknowledgements}

\bibliographystyle{aa}
\bibliography{refs}

\begin{appendix}

\FloatBarrier
\onecolumn

\section{Test 2 results}

\begin{figure}[h!]
    \centering
    \begin{subfigure}[b]{0.48\textwidth}
        \centering
        \includegraphics[width=\textwidth]{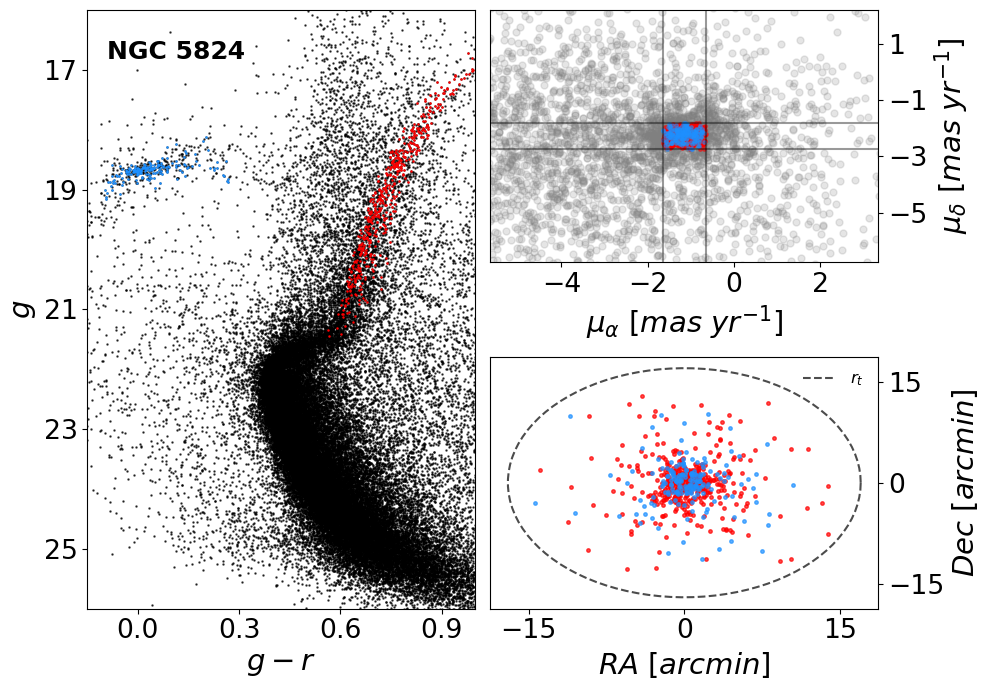}
    \end{subfigure}
    \hfill
    \begin{subfigure}[b]{0.48\textwidth}
        \centering
        \includegraphics[width=\textwidth]{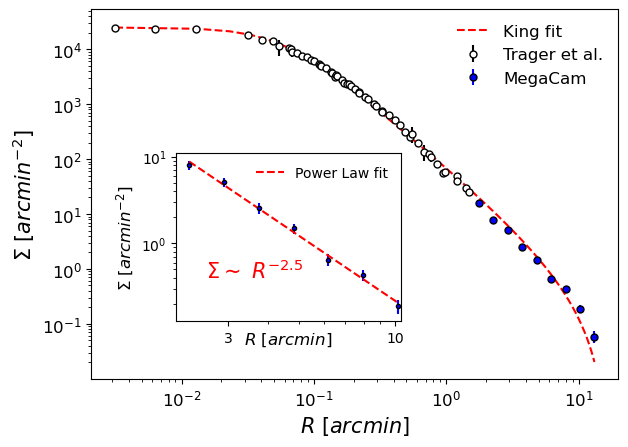}
    \end{subfigure}

    \vspace{0.5em}
    \begin{subfigure}[b]{0.48\textwidth}
        \centering
        \includegraphics[width=\textwidth]{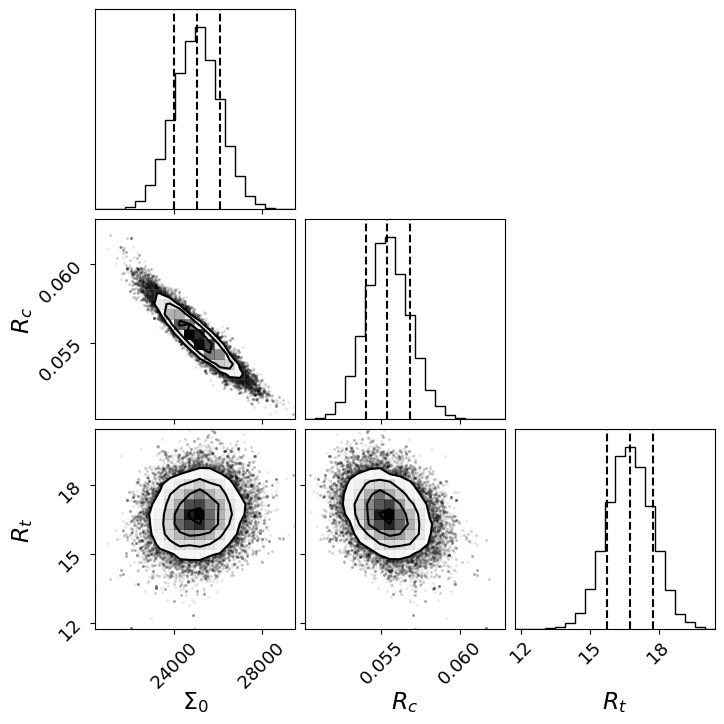}
    \end{subfigure}
    \hfill
    \begin{subfigure}[b]{0.48\textwidth}
        \centering
        \includegraphics[width=\textwidth]{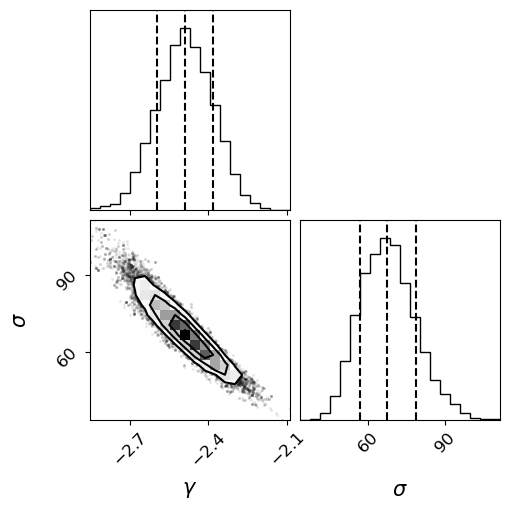}
    \end{subfigure}

    \caption{Same as Fig. \ref{fig:test1_results} but for Test 2, which uses MegaCam photometry and \textit{Gaia} proper motions. The top-left panel includes a proper motion diagram where gray dots represent the full dataset with PM measurements, and red and blue dots correspond to the RGB and BHB stars candidates, respectively, which are also shown in the CMD and star-count map.}
    \label{fig:test2_results}
\end{figure}
\newpage

\FloatBarrier
\section{Test 3 results}

\begin{figure}[h!]
    \centering
    \begin{subfigure}[b]{0.48\textwidth}
        \centering
        \includegraphics[width=\textwidth]{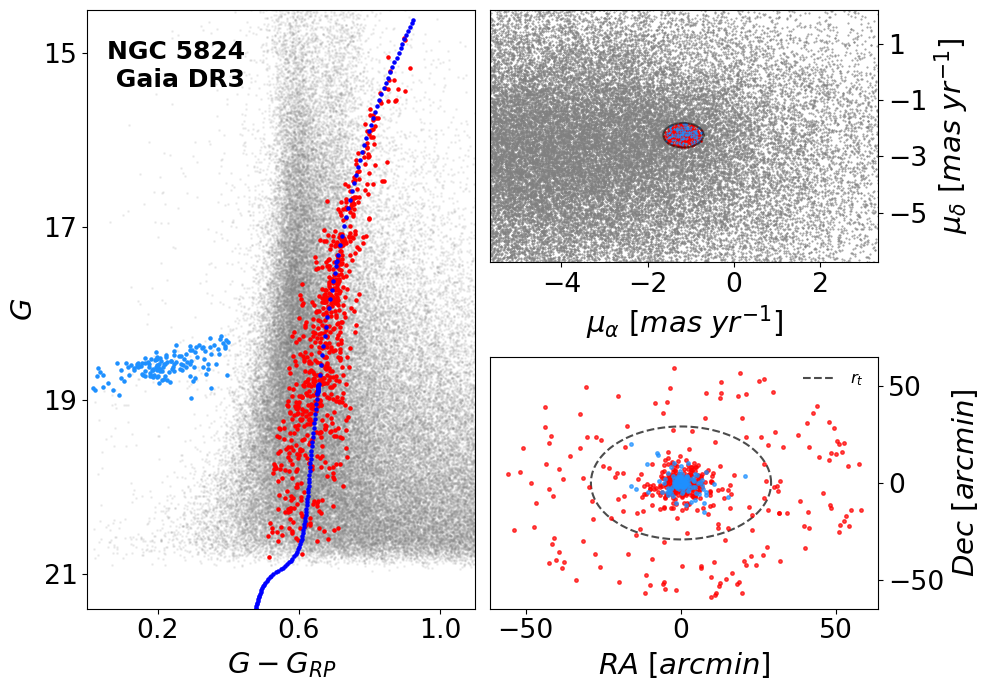}
    \end{subfigure}
    \hfill
    \begin{subfigure}[b]{0.48\textwidth}
        \centering
        \includegraphics[width=\textwidth]{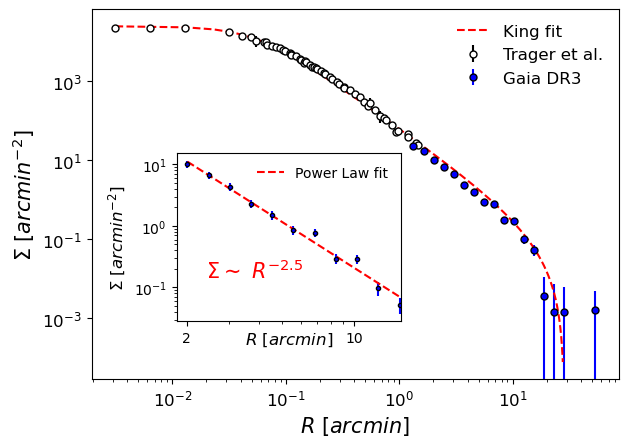}
    \end{subfigure}

    \vspace{0.5em}
    \begin{subfigure}[b]{0.48\textwidth}
        \centering
        \includegraphics[width=\textwidth]{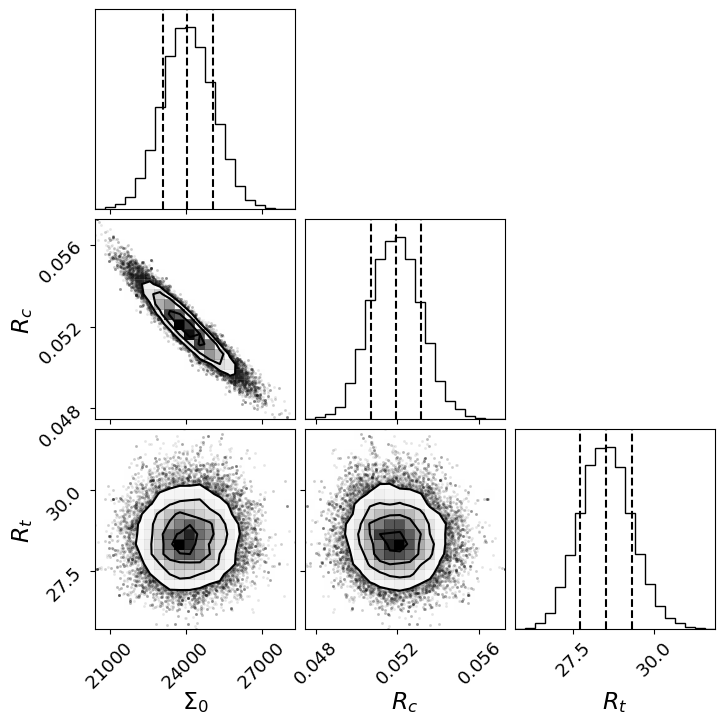}
    \end{subfigure}
    \hfill
    \begin{subfigure}[b]{0.48\textwidth}
        \centering
        \includegraphics[width=\textwidth]{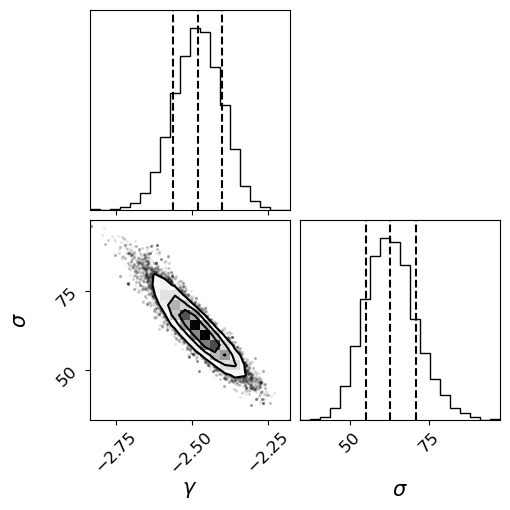}
    \end{subfigure}
    
    \caption{Same as Fig. \ref{fig:test2_results} but for Test 3, which uses \textit{Gaia} DR3 photometry and proper motion.}
    \label{fig:test3_results}
\end{figure}
\newpage

\FloatBarrier
\section{Test 4 results}

\begin{figure}[h!]
    \centering
    \begin{subfigure}[b]{0.48\textwidth}
        \centering
        \includegraphics[width=\textwidth]{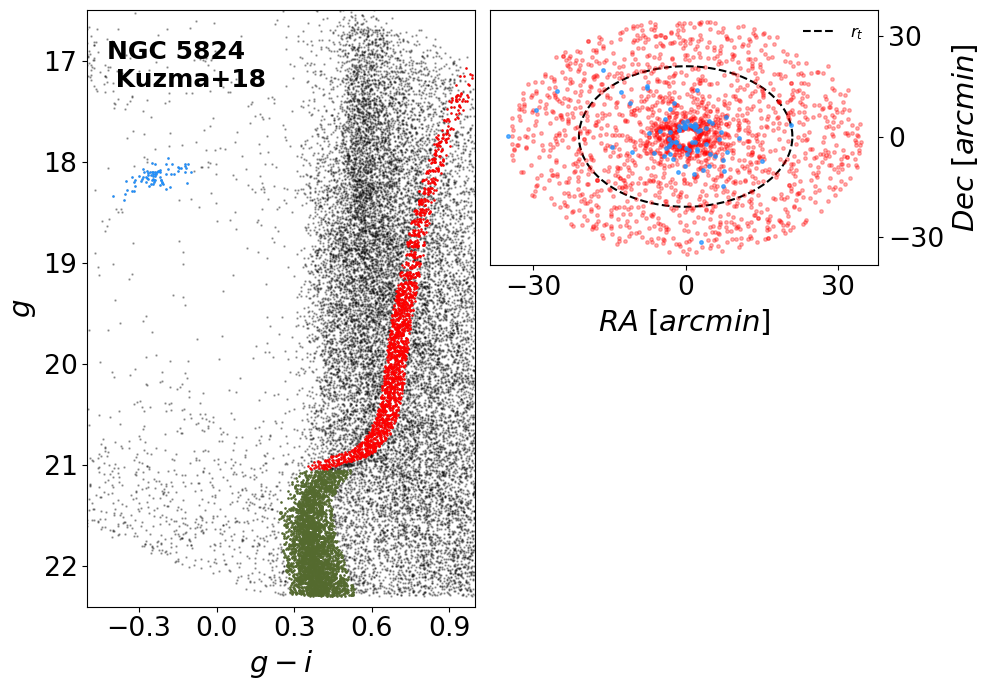}
    \end{subfigure}
    \hfill
    \begin{subfigure}[b]{0.48\textwidth}
        \centering
        \includegraphics[width=\textwidth]{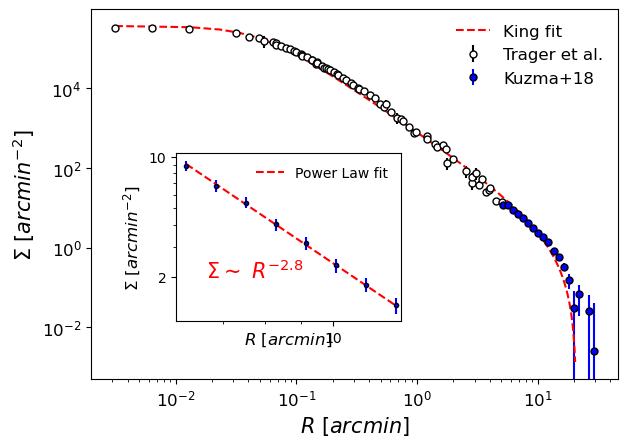}
    \end{subfigure}

    \vspace{0.5em}
    \begin{subfigure}[b]{0.48\textwidth}
        \centering
        \includegraphics[width=\textwidth]{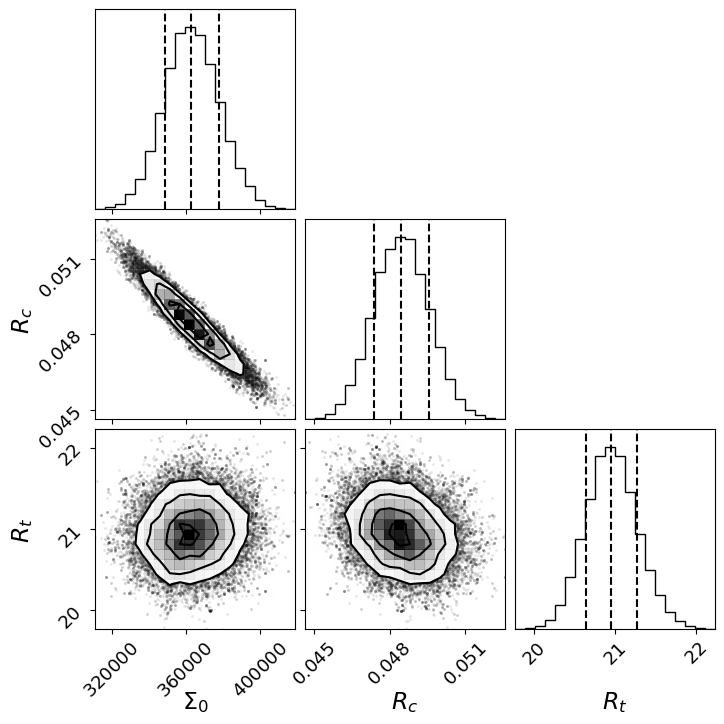}
    \end{subfigure}
    \hfill
    \begin{subfigure}[b]{0.48\textwidth}
        \centering
        \includegraphics[width=\textwidth]{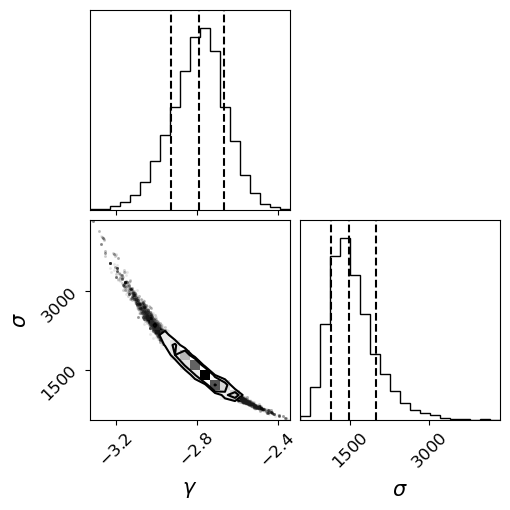}
    \end{subfigure}

    \caption{Same as Fig. \ref{fig:test1_results} but for Test 4, which uses DECam photometry from \cite{Kuzma_2018}.}
    \label{fig:test4_results}
\end{figure}
\newpage

\FloatBarrier
\section{NGC\,2419 results}

\begin{figure}[h!]
    \centering
    \begin{subfigure}[b]{0.48\textwidth}
        \centering
        \includegraphics[width=\textwidth]{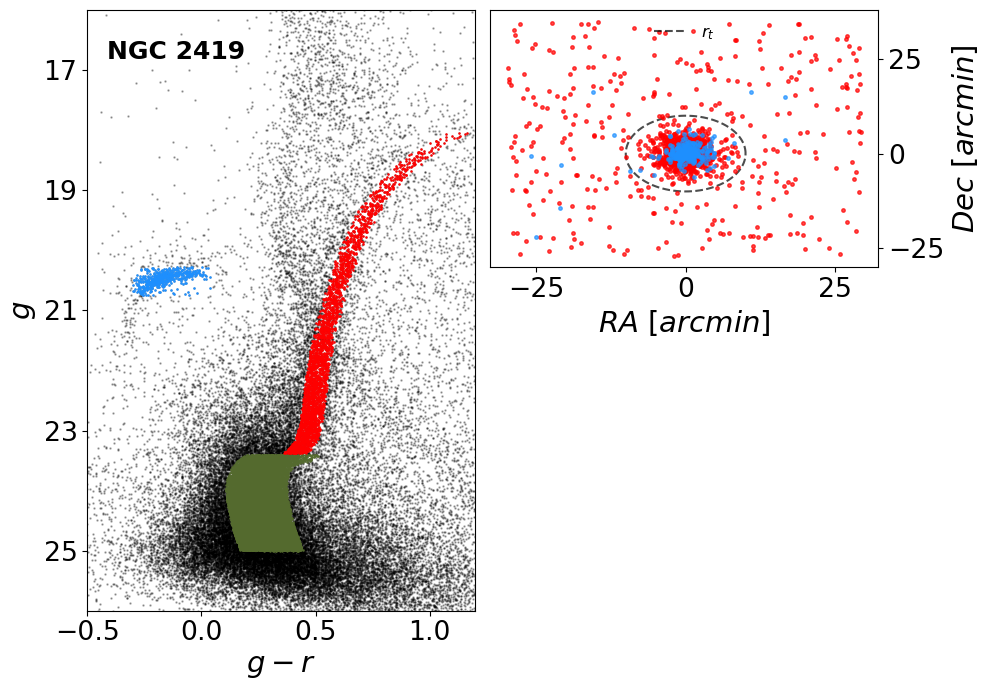}
    \end{subfigure}
    \hfill
    \begin{subfigure}[b]{0.48\textwidth}
        \centering
        \includegraphics[width=\textwidth]{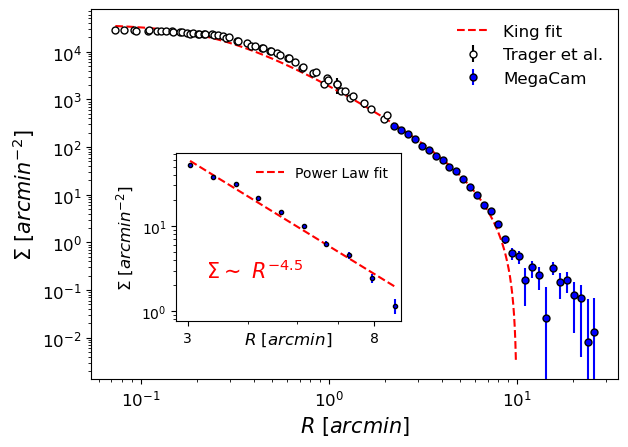}
    \end{subfigure}

    \vspace{0.5em}
    \begin{subfigure}[b]{0.48\textwidth}
        \centering
        \includegraphics[width=\textwidth]{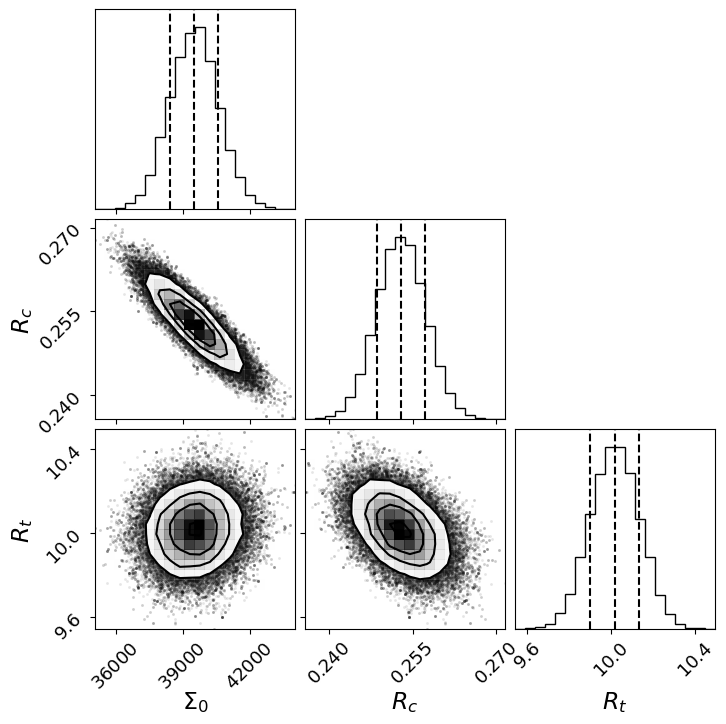}
    \end{subfigure}
    \hfill
    \begin{subfigure}[b]{0.48\textwidth}
        \centering
        \includegraphics[width=\textwidth]{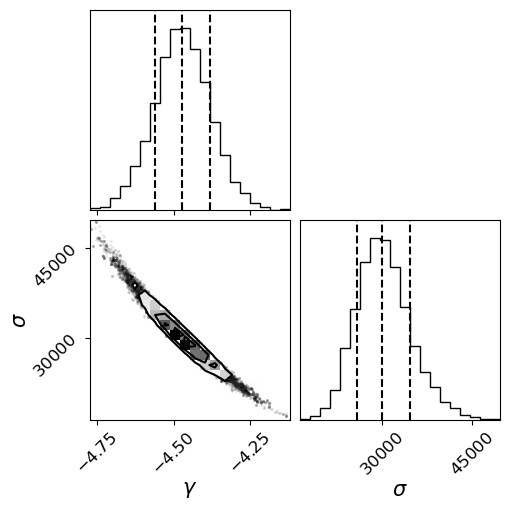}
    \end{subfigure}
    
    \caption{Same as Fig. \ref{fig:test1_results} but for NGC\,2419. For the King profile fitting, we used density values up to the background level, corresponding to $\approx 9'$ from the cluster center. This is motivated by the fact that density values below the background level exhibit high signal-to-noise ratios, causing the MCMC algorithm to misinterpret and include them in the fit, leading to an inadequate result.}
    \label{fig:ngc2419_results}
\end{figure}
\twocolumn

\end{appendix}
\end{document}